\documentclass[12pt,letterpaper]{iopart}
\usepackage{natbib}

\newcommand{\arcsec}{^{\prime\prime}}
\newcommand{\farcs}{.\!\!^{\prime\prime}}
\newcommand{\pasp}{\textit{Pub.\ Astron.\ Soc.\ Pacific}}
\newcommand{\mnras}{\textit{Mon.\ Not.\ Royal Astron.\ Soc.}}
\newcommand{\aap}{\textit{Astron.\ \& Astrophys.}}
\newcommand{\apj}{\textit{Astrophys.\ J.}}
\newcommand{\apjl}{\textit{Astrophys.\ J.\ Lett.}}
\newcommand{\aj}{\textit{Astron.\ J.}}
\usepackage{graphicx}
\begin{document}

\title[An integral-field spectroscopic strong lens survey]{An integral-field spectroscopic strong lens survey$^{\dag}$}
\address{$^{\dag}$ This paper includes data gathered with the 6.5 meter Magellan Telescopes located at Las Campanas Observatory, Chile and data gathered with
the 8.1 meter Gemini Telescope located on Mauna Kea, Hawai`i.}
\author{Adam S Bolton$^1$ and Scott Burles$^2$}
\address{$^1$ Harvard-Smithsonian Center for Astrophysics, 60 Garden St. MS-20,
Cambridge, MA 02138 USA}
\address{$^2$ Department of Physics and Kavli Institute, Massachusetts Institute of Technology, 77 Massachusetts Ave., Cambridge, MA 02139 USA}
\eads{\mailto{abolton@cfa.harvard.edu}, \mailto{burles@mit.edu}}
\vspace*{12pt}

\submitto{``Gravitational Lensing'' Focus Issue of the New Journal of Physics (invited)}

\begin{abstract}
We present the observational results of a survey for
strong gravitational lens systems consisting
of extended emission-line galaxies lensed by
intervening early-type galaxies, conducted using integral
field units (IFUs) of the Magellan IMACS and
Gemini GMOS-N spectrographs.  These data are highly
valuable for corroborating the lensing interpretation of
Hubble Space Telescope imaging data.  We show that in many cases,
ground-based IFU spectroscopy is in fact competitive with
space-based imaging for the measurement of the mass model parameters
of the lensing galaxy.  We demonstrate a novel technique of
three-dimensional gravitational lens modeling for a single
lens system with a resolved lensed rotation curve.  We also
describe the details of our custom IFU data analysis software,
which performs optimal multi-fiber extraction, relative and
absolute wavelength calibration to a few hundredths
of a pixel RMS, and nearly Poisson-limited sky subtraction.
\end{abstract}

\maketitle

\section{Introduction}

In the pursuit of strong lensing science, spatial resolution
is of paramount importance, and therefore high-resolution imaging
with the \textsl{Hubble Space Telescope} (\textsl{HST}) currently sets
the gold standard at optical wavelengths.  
This paper presents a different observational approach: a
spatially resolved spectroscopic survey for strong galaxy-galaxy
gravitational lenses using the optical
integral-field units (IFUs)
of the Gemini GMOS-N and Magellan IMACS spectrographs.
A significant goal of this paper is to demonstrate
the potential of this technique for the confirmation and study
of strong lenses.  We will show that,
in comparison with \textsl{HST} imaging, ground-based
IFU spectroscopy is always complementary, often competitive,
and occasionally superior.  These favorable comparisons are
all due to the combination of large collecting aperture with
simultaneous spatial and spectral resolution, currently only available
at ground-based observatories.  Specifically, we show that
IFU observations can be valuable for corroborating the lensing
interpretation of high-resolution imaging data by using
the third (wavelength) dimension to cleanly separate the
narrow-band images of two separate galaxies along a single
line of sight.  We show that in many cases, ground-based
IFU data alone can provide competitive constraints on the
mass model of the lensing galaxy as compared to \textsl{HST}
data.  Finally, we show that IFU data can be superior
to space-based imaging in cases where the lensed emission-line
flux gives a clearer picture than the lensed continuum, or when
the presence of resolved velocity structure in the lensed
galaxy allows a three-dimensional approach to lens modeling.

Numerous previous studies have used the unique
three-dimensional capability of high spatial sampling IFUs for
galaxy-scale strong lensing science.
The majority of these works have used
IFU spectroscopy to measure differential substructure
lensing effects on the
continuum and emission-line regions of lensed quasars
\citep{mediavilla_98, wisotzki_03, met_mous_04,
motta_04, wayth_05, lamer_06, sugai_07}.
IFUs have also been used to resolve the kinematic structure of high-redshift
galaxies lensed into magnified arcs by intervening galaxy clusters
\citep{swinbank_03, swinbank_06, swinbank_07}.  The results we
present here represent the first application of IFU spectroscopy
to the confirmation and modeling of spectroscopically selected
strong gravitational lens candidates consisting of a superposition
of two galaxies along a single line of sight.

A further goal of this paper is to present the details
of our custom IFU data calibration and analysis software.
This software (described in \ref{ifu_red}) performs optimal
extraction, relative and absolute wavelength calibration
to a few hundredths of a pixel RMS, and nearly Poisson-limited
sky subtraction.

\section{Sample and observations}

The lens candidates observed for this study were selected
from the spatially unresolved spectroscopic database of the
Sloan Digital Sky Survey (SDSS; \citealt{york_sdss}) using
the technique described in \citet{bolton_speclens}.
They are identified through the presence of an emission-line redshift
significantly higher than the absorption-line redshift within
the same spectrum, as observed through the 3$\arcsec$-diameter
SDSS spectroscopic fiber aperture.  By design of the selection
process, the candidates consist of a close angular superposition
of a bright foreground early-type galaxy with a faint
background star-forming galaxy.  If the (initially unknown)
impact parameter is small enough, the background galaxy
will be multiply imaged by the gravity of the intervening early-type,
furnishing a powerful probe of the distribution of mass
in the latter.  Many such candidates, including all the confirmed
lenses presented in this paper, have been successfully observed
with the Advanced Camera for Surveys (ACS) aboard the
\textsl{Hubble Space Telescope} (\textsl{HST})
by the Sloan Lens ACS (SLACS) Survey
\citep{bolton_1402, slacs1, slacs2, slacs3, slacs4, slacs5}.

The observations for this work were obtained with
with IFUs built by the University
of Durham Astronomical Instrumentation Group for the
The Inamori Magellan Areal Camera and Spectrograph
\citep[IMACS;][]{bigelow_imacs} and the Gemini-North Multi-Object
Spectrograph \citep[GMOS-N;][]{hook_gmos}.
The data were collected during 2004 March--May (GMOS-N)
and 2004 September (IMACS).  IMACS data were also collected during
2003 November and 2004 March, but observing and instrument
conditions were too poor for scientific use.
Table~\ref{infotab} gives information on the candidate
systems for which data is presented in this paper.

\begin{table}
\caption{\label{infotab}
Information for double-redshift systems with
IFU data presented in this paper.  SDSS spectroscopic
ID consists of plate number, modified Julian date,
and fiber number. The columns $z_{\mathrm{FG}}$ and $z_{\mathrm{BG}}$
respectively contain foreground and background
redshifts measured for each system from SDSS spectroscopy.
The magnitudes $r$ are extinction-corrected
\citep{sfd_dust} deVaucouleurs model values computed
from SDSS imaging data, and are dominated by the flux
of the foreground galaxy.}
\begin{indented}
\item[]\begin{tabular}{@{}llllll}
\br
Name & SDSS spec.\ ID & J2000 RA \& Dec. &
$z_{\mathrm{FG}}$ & $z_{\mathrm{BG}}$ & $r$ \\
\mr
  SDSSJ0037$-$0942 & 0655-52162-392 & 00:37:53.21$-$09:42:20.1 & 0.1954 & 0.6322 & 16.8 \\
  SDSSJ0044$+$0113 & 0393-51794-456 & 00:44:02.90$+$01:13:12.5 & 0.1196 & 0.1967 & 16.2 \\
  SDSSJ0737$+$3216 & 0541-51959-145 & 07:37:28.44$+$32:16:18.6 & 0.3223 & 0.5812 & 17.9 \\
  SDSSJ0928$+$4400 & 0870-52325-465 & 09:28:57.33$+$44:00:59.1 & 0.2909 & 0.4538 & 18.0 \\
  SDSSJ0956$+$5100 & 0902-52409-068 & 09:56:29.78$+$51:00:06.3 & 0.2405 & 0.4700 & 17.2 \\
  SDSSJ1029$+$6115 & 0772-52375-140 & 10:29:27.53$+$61:15:05.0 & 0.1573 & 0.2512 & 16.5 \\
  SDSSJ1128$+$5835 & 0951-52398-036 & 11:28:37.77$+$58:35:26.8 & 0.3809 & 0.5466 & 18.3 \\
  SDSSJ1155$+$6237 & 0777-52320-501 & 11:55:10.09$+$62:37:22.1 & 0.3751 & 0.6690 & 18.2 \\
  SDSSJ1259$+$6134 & 0783-52325-279 & 12:59:19.06$+$61:34:08.4 & 0.2333 & 0.4488 & 17.5 \\
  SDSSJ1402$+$6321 & 0605-52353-503 & 14:02:28.22$+$63:21:33.3 & 0.2046 & 0.4814 & 17.0 \\
  SDSSJ1416$+$5136 & 1045-52725-464 & 14:16:22.33$+$51:36:30.2 & 0.2987 & 0.8115 & 18.1 \\
  SDSSJ1521$+$5805 & 0615-52347-311 & 15:21:23.87$+$58:05:50.6 & 0.2042 & 0.4857 & 17.4 \\
  SDSSJ1630$+$4520 & 0626-52057-518 & 16:30:28.15$+$45:20:36.2 & 0.2479 & 0.7933 & 17.4 \\
  SDSSJ1702$+$3320 & 0973-52426-464 & 17:02:16.76$+$33:20:44.7 & 0.1784 & 0.4357 & 16.9 \\
  SDSSJ2238$-$0754 & 0722-52224-442 & 22:38:40.20$-$07:54:56.0 & 0.1371 & 0.7126 & 16.8 \\
  SDSSJ2302$-$0840 & 0725-52258-463 & 23:02:20.17$-$08:40:49.4 & 0.0901 & 0.2223 & 16.9 \\
  SDSSJ2321$-$0939 & 0645-52203-517 & 23:21:20.93$-$09:39:10.3 & 0.0819 & 0.5324 & 15.2 \\
\br
\end{tabular}
\end{indented}
\end{table}

The IMACS IFU \citep{schmoll_imacs_ifu} observes
two $5\arcsec \times 7 \arcsec$ fields of view
in the focal
plane, separated by
roughly one arc-minute: one FOV for the object and one from which
to estimate the sky background.
The fields are sampled by a close-packed hexagonal array of
lenslets which subtend $0\farcs 2$ from side to side, for a total
of 2000 lenslets between the two fields.  The lenslets feed
the light to optical fibers, which reformat the fields
via a defined field-mapping into a
one-dimensional array of output lenslets (a ``pseudo-slit'')
for dispersion.  This is accomplished within the space
of a narrow cartridge that occupies the width
of three adjacent mask
slots in the slit-mask server, which inserts and removes the
IFU in the same manner as a simple mask.  Thus the IFU behaves
like a slit-mask with two $5\arcsec \times 7 \arcsec$ slits on
the input side and one long slit on the output.
The IMACS IFU can be used with either of the two IMACS
cameras: $f/4$ (``long'') or $f/2$ (``short'').
All IMACS data presented in this paper were obtained using the $f/2$ camera,
which uses grisms for dispersion.
The GMOS-N IFU \citep{jas_gmos_ifu, murray_gmos_ifu}
operates in an identical manner to the IMACS IFU,
with a few notable exceptions.  First, while the object field
is $5\arcsec \times 7 \arcsec$ (1000 lenslets) as in IMACS, the
background field is half this size: $5\arcsec \times 3\farcs 5$
(500 lenslets).  Second, the 1500 total fibers are reformatted
not into one single pseudo-slit, but rather into two parallel
pseudo-slits separated by approximately 3200 pixels on the detector.
In the ``two-slit'' mode employed in the current work,
broad-band filters must be
used to limit the wavelength domain of the individual
pseudo-slits so as to prevent overlapping of spectra.
Table~\ref{modetab} lists the various unique spectrograph
configurations used in the current work, along with their
general characteristics.

\begin{table}
\caption{\label{modetab}
Unique Gemini GMOS-N and Magellan-IMACS IFU observing modes used to collect
the data presented in this paper.  Three additional targets were observed
using GMOS-N in the $g$ band, but the data were not successfully reduced
due to problems with tracing, flat-fielding, and wavelength calibration.
All GMOS-N IFU observations were binned $2\times$ in the dispersion
direction.}
\begin{indented}
\item[]\begin{tabular}{@{}llllll}
\br
Short &  ~   &     ~     & Blocking & ~     & $R = $ \\
name &  Mode & Disperser & filter   & Range & $\lambda / \Delta \lambda$ \\
\mr
IMACS  & f/2 camera & 300 l/mm grism & none      & 4000--9000\AA & $\sim 2000$ \\
GMOS-r & Two-slit     & R600 grating   & Sloan $r$ & 5500--7000\AA & $\sim 4000$ \\
GMOS-i & Two-slit     & R600 grating   & Sloan $i$ & 7000--8500\AA & $\sim 4000$ \\
\br
\end{tabular}
\end{indented}
\end{table}

\section{Narrow-band image construction}

The three-dimensional data provided by IFUs affords the
opportunity to construct narrow-band images of any bandwidth and at
any wavelength within the data cube.  Furthermore,
emission-line regions of the spectrum may be decomposed into
continuum and emission-line components through suitable profile modeling.
To a very good approximation---largely as a consequence of
our initial selection bias---continuum emission seen in the IFU spectra of
our lens candidates can be attributed to
the foreground galaxy.  Similarly, high-redshift emission
lines is of course entirely due to the background
galaxies.  Thus decomposition into continuum and emission
line components is approximately equivalent to decomposition into
foreground and background galaxy images.

Our strategy for continuum/emission-line decomposition is as follows.
First, we select a small ($\pm \sim 10$ \AA\ ) wavelength range about
the wavelength of a background emission line detected in the
SDSS spectroscopy.  We parameterize the emission-line component as
a Gaussian function of wavelength, and the continuum as either a
constant or linear function of wavelength, which are
sufficient for our purposes over these small wavelength ranges.
The central wavelength and width of the Gaussian term, as well as the
dimensionless slope of the continuum, are treated as
global model parameters applying to all fiber spectra.  For a given
trial set of these parameters, a basis is constructed
and the amplitudes of the continuum and
emission-line components are fitted with a linear
least-squares fit to the IFU data over
the selected wavelength range.  This linear fit
is wrapped within a non-linear optimization of the
global emission-line and continuum shape parameters
that minimizes the overall $\chi^2$ using the MPFIT IDL
implementation of the
Levenberg-Marquardt algorithm.  For the case of
[O\textsc{ii}]~3727 emission, a double Gaussian line profile is used, with
the line spacing fixed to its known rest-frame value of $2.78$ \AA\
and the relative strength of the two line components fitted as an
additional global parameter.   In several systems, slight velocity
shifts are detected in the background emission lines, and these
are modeled with an additional Gaussian-derivative amplitude that
approximates the small velocity shifts.  One system shows an
appreciably resolved rotation curve, which is treated separately
as described below.

\begin{figure}
\centerline{\scalebox{0.38}{\includegraphics{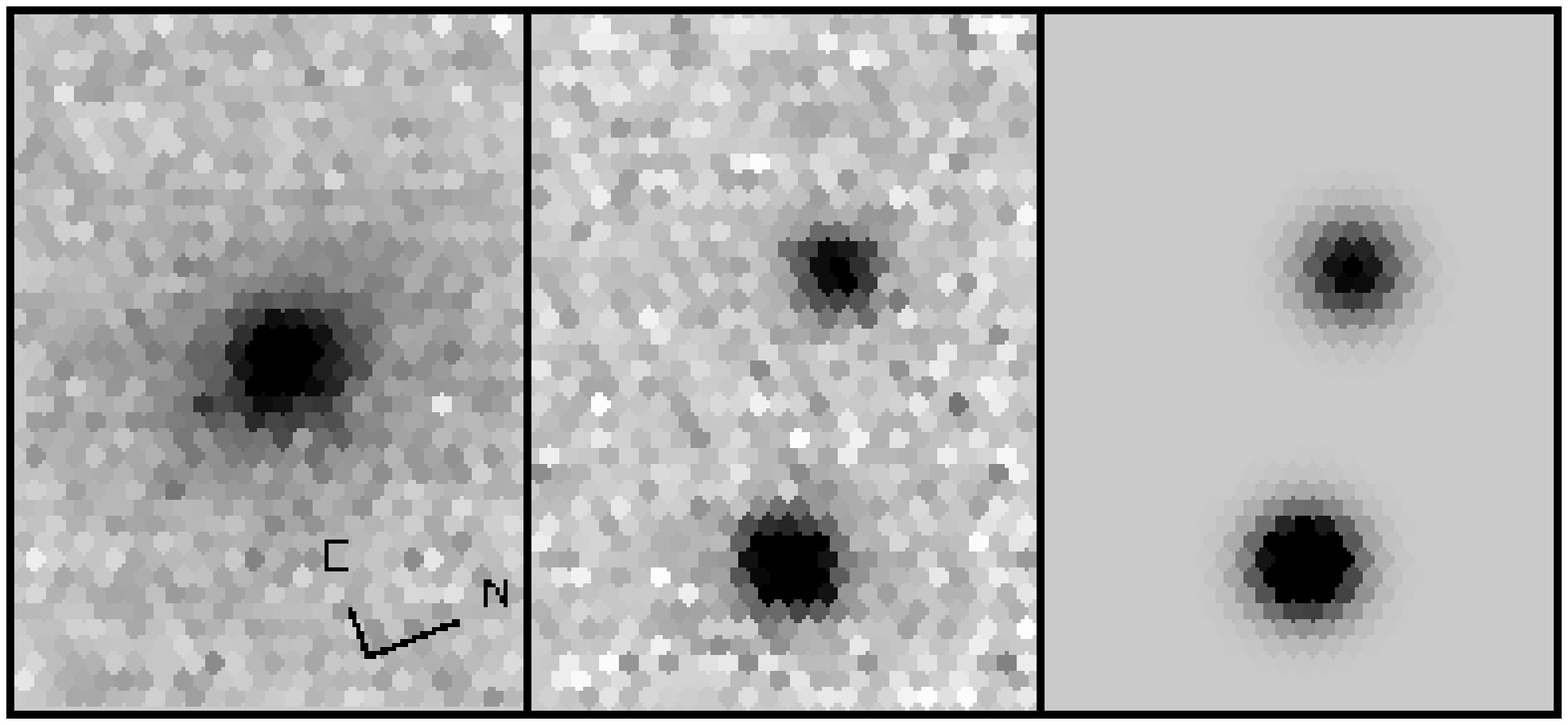}\hspace*{12pt}\includegraphics{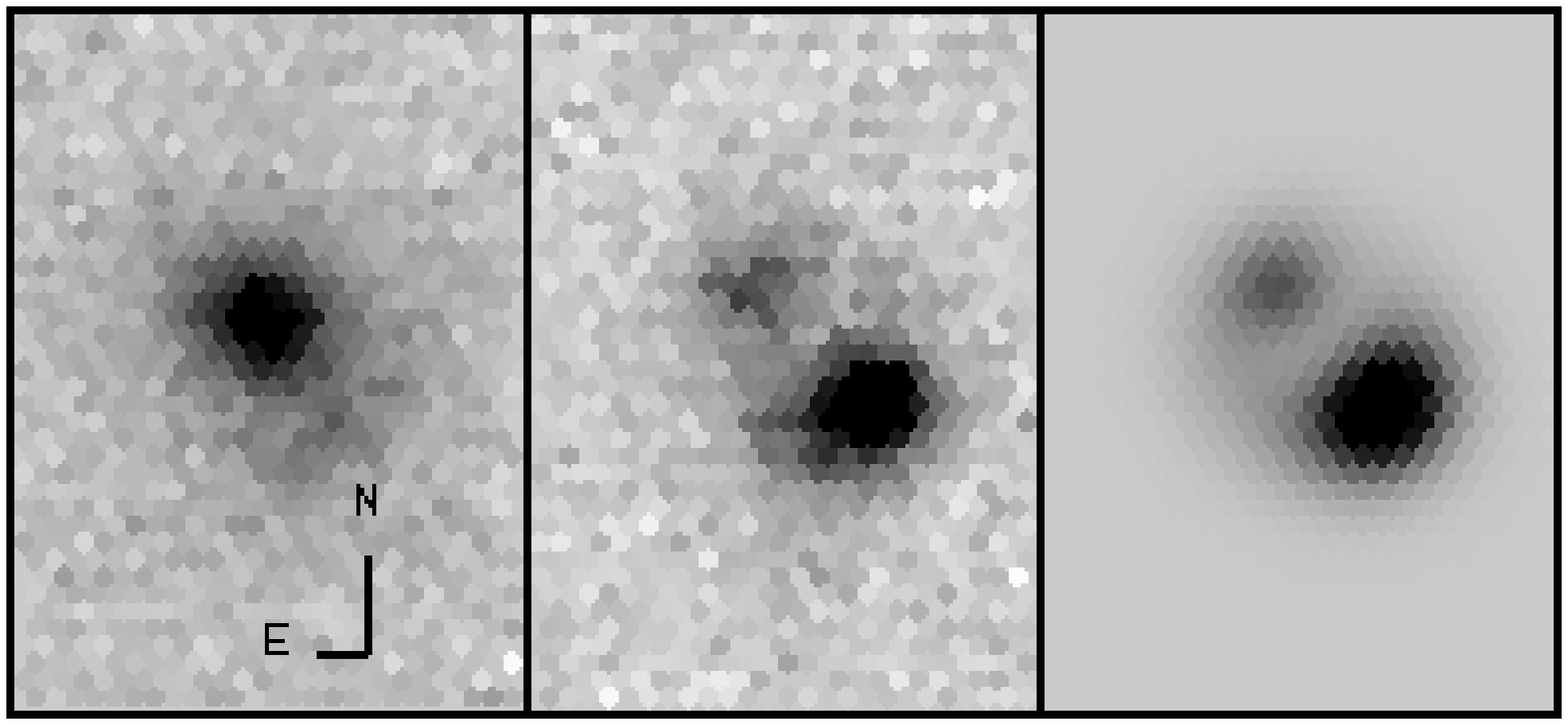}}}
\vspace*{5pt}
\centerline{\scalebox{0.38}{\includegraphics{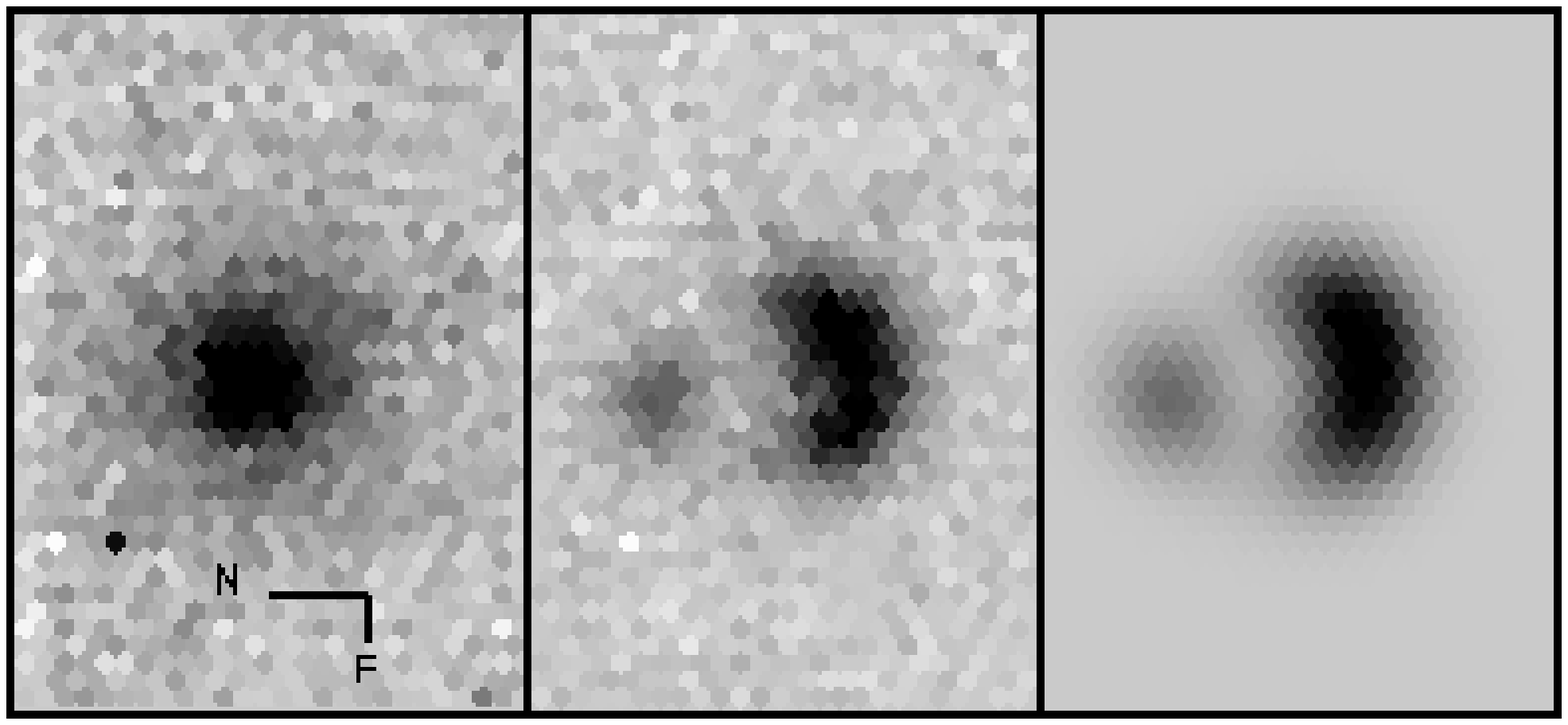}\hspace*{12pt}\includegraphics{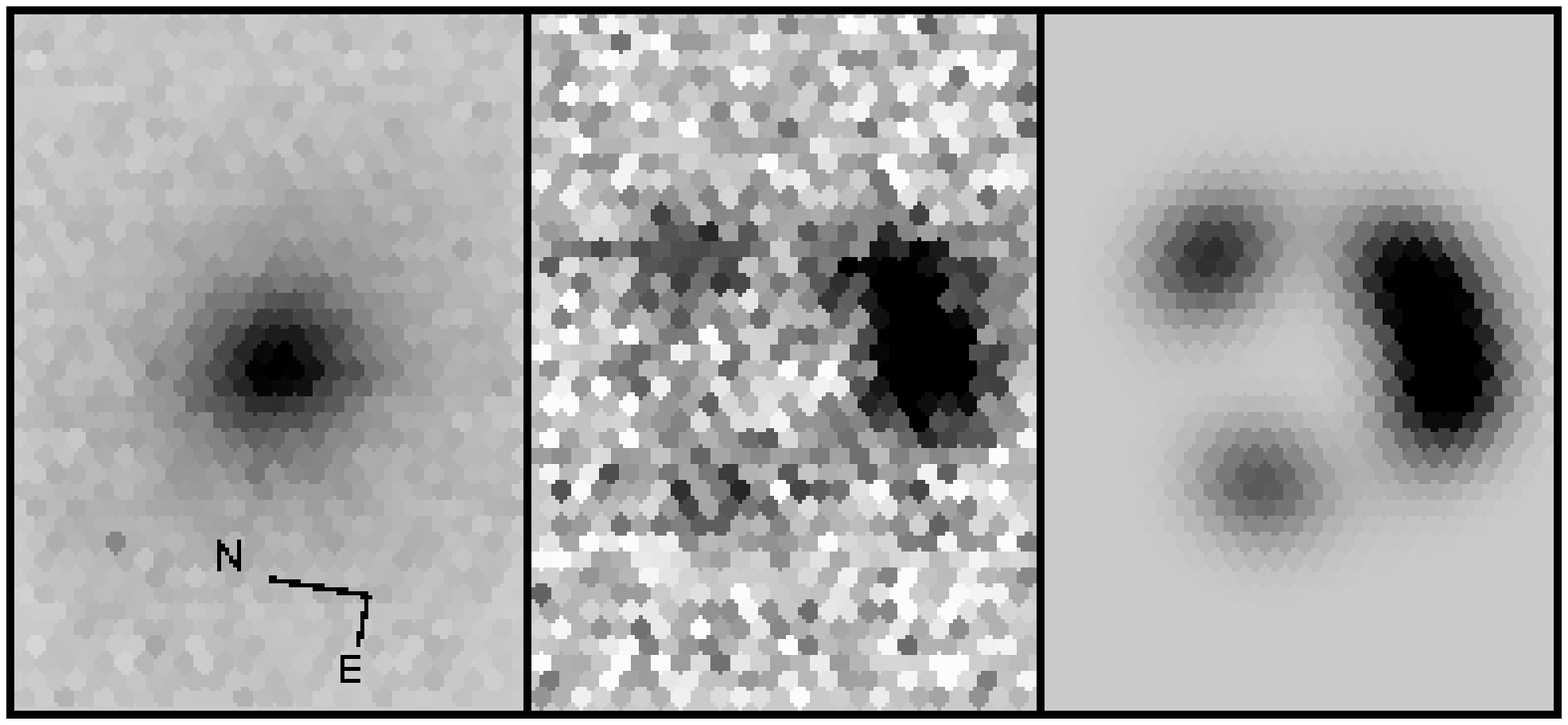}}}
\vspace*{5pt}
\centerline{\scalebox{0.38}{\includegraphics{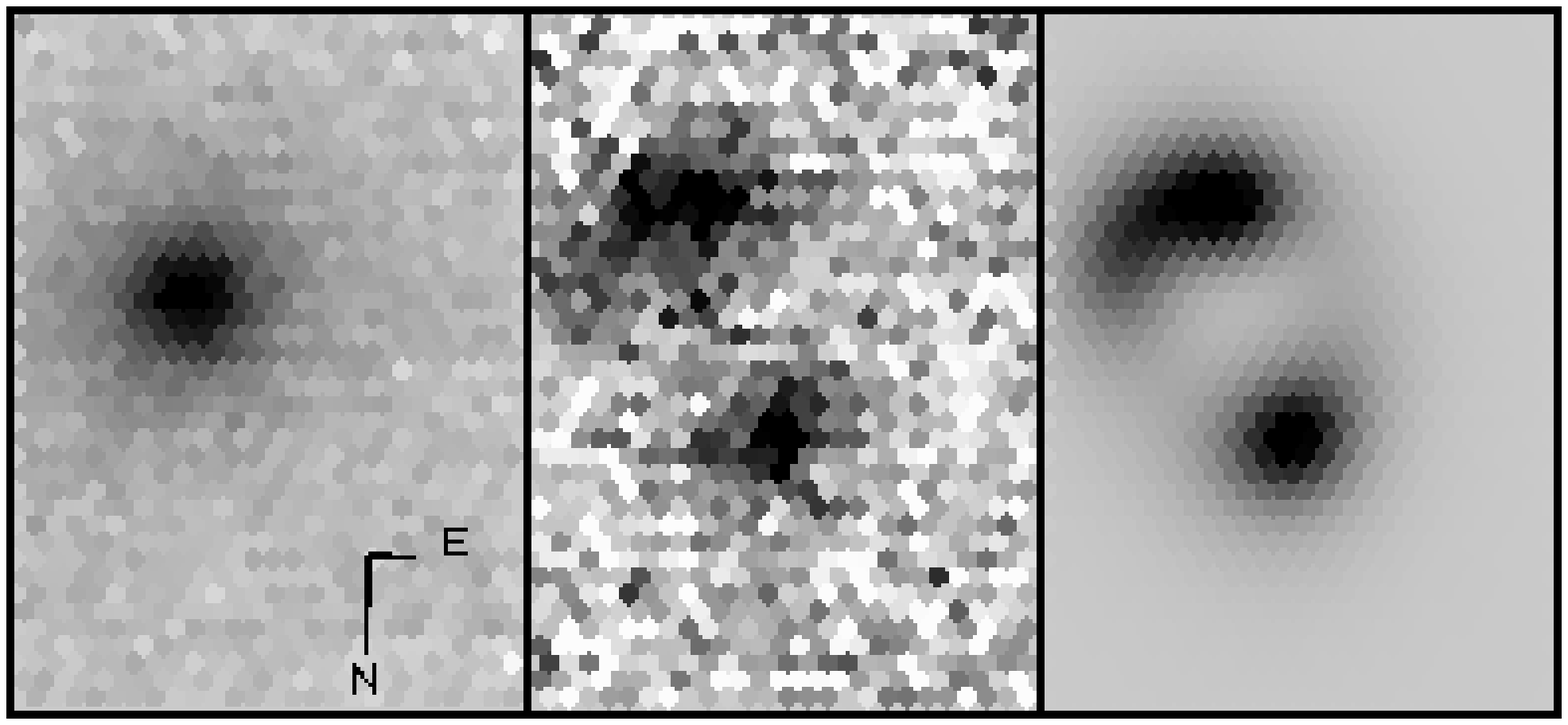}\hspace*{12pt}\includegraphics{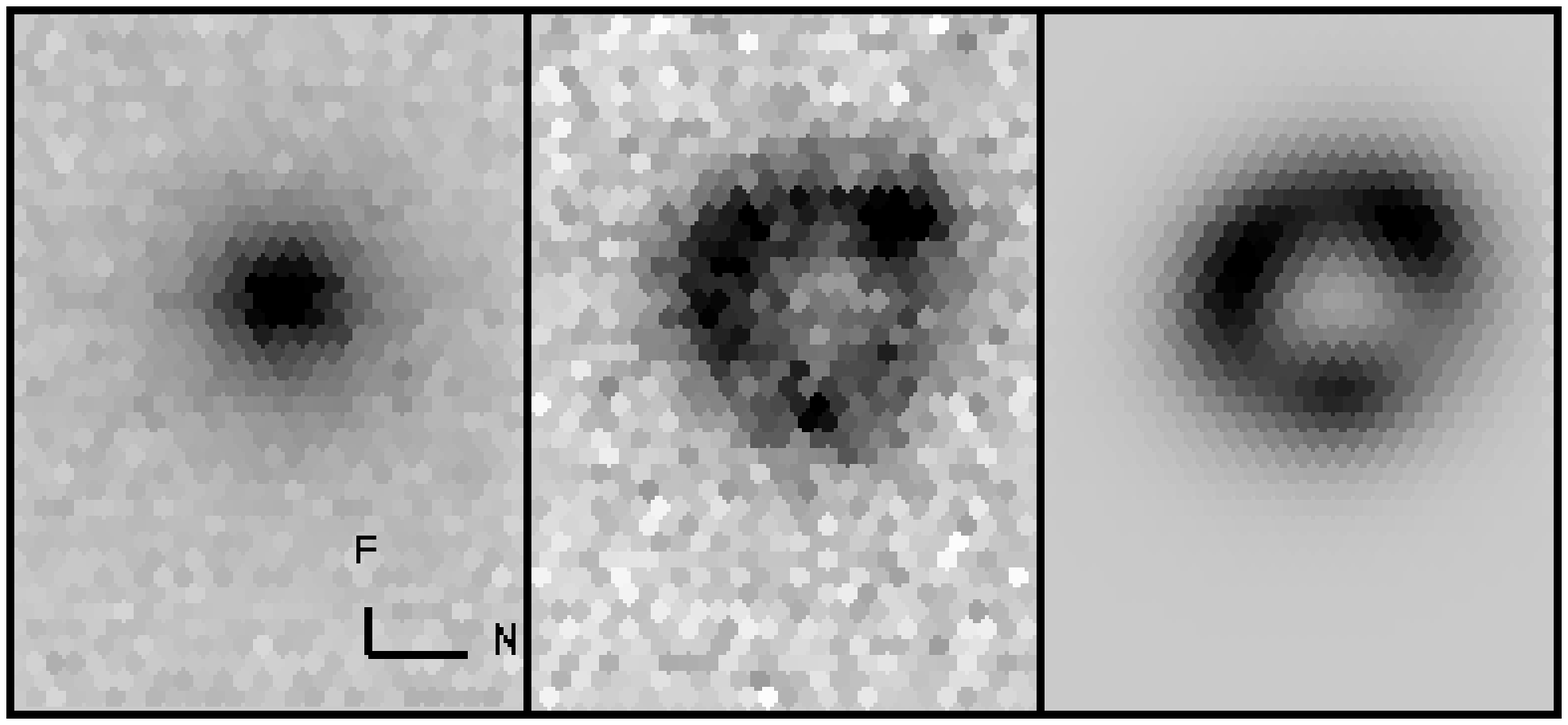}}}
\vspace*{5pt}
\centerline{\scalebox{0.38}{\includegraphics{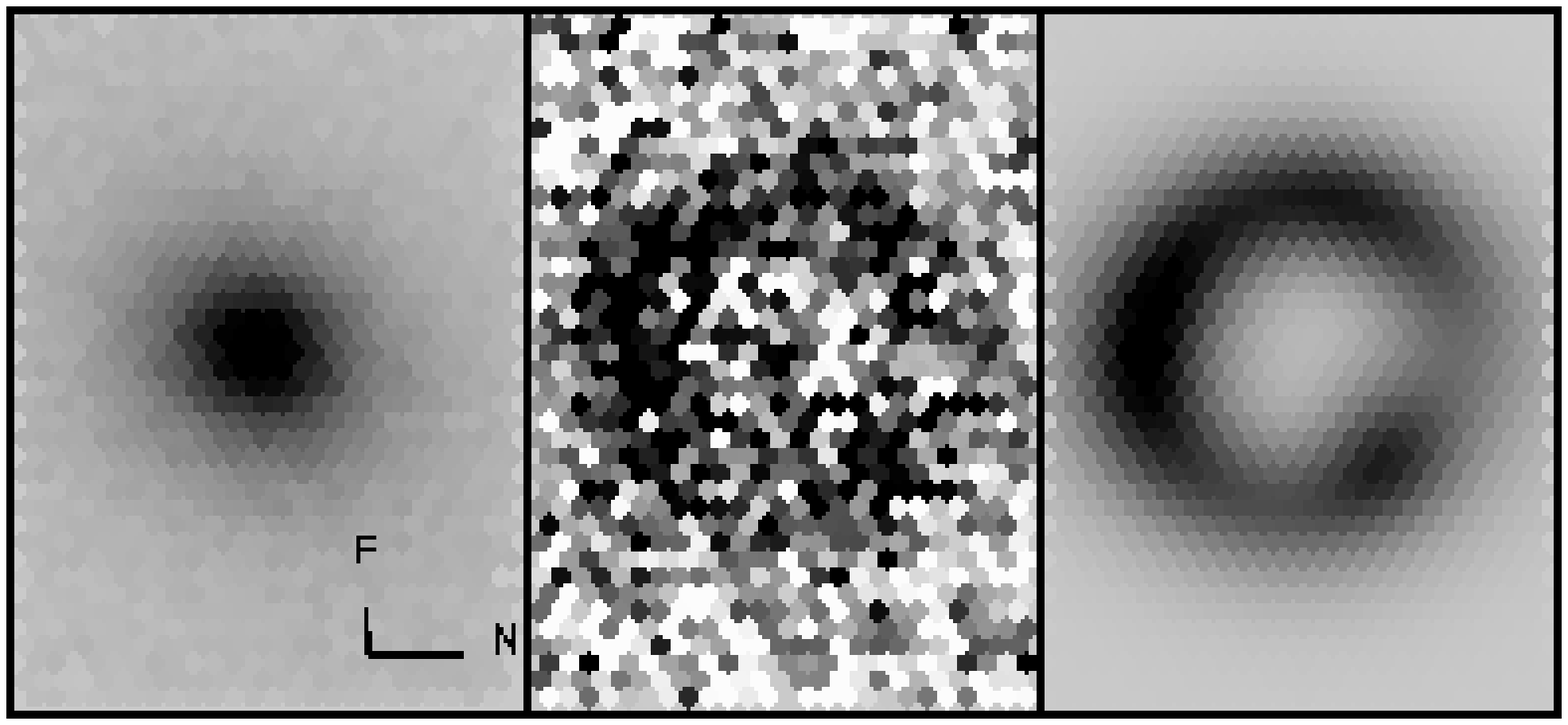}}}
\caption{\label{lensmodels}
Lens systems that admit successful IFU gravitational-lens modeling.
There are three panels for each system: narrow-band continuum level (left),
narrow-band emission-line level (center), and model emission-line level as
reproduced by the best-fit gravitational lens model (right).  All panels
are $5 \arcsec \times 7 \arcsec$.  Gray-scale is linear, and varies from
system to system and between continuum and emission-line panels for
visual ease.  Individual gray-scale limits are given in Table~\ref{datatab}.
Black corresponds to values at the limit and above, and white corresponds to
values at $-0.25$ times the limit and below.
In reading order from upper left, the individual systems are:
J0037, J0044, J0737, J1402, J2238, J2302, and J2231.}
\end{figure}

\begin{figure}
\centerline{\scalebox{0.38}{\includegraphics{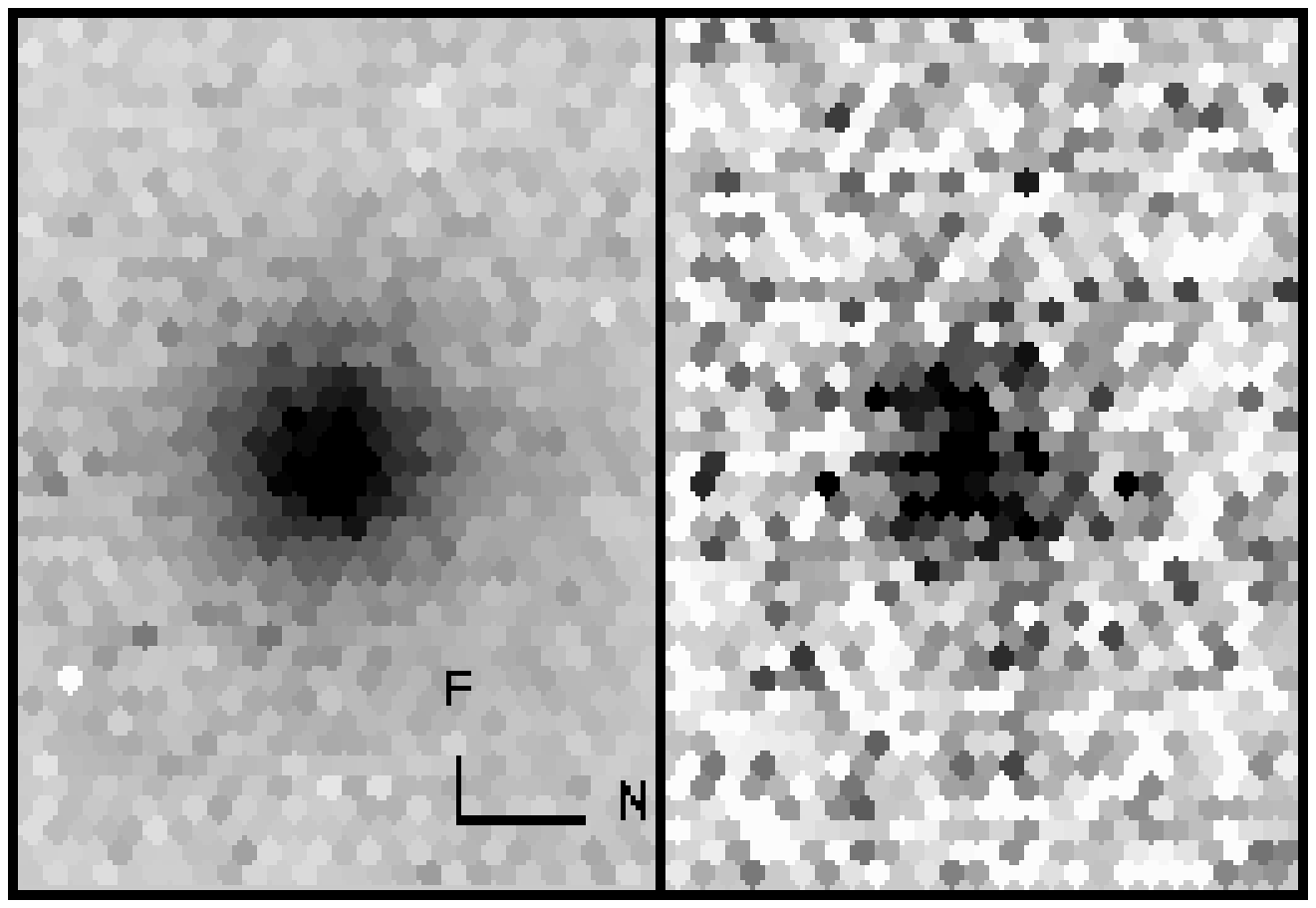}\hspace*{12pt}\includegraphics{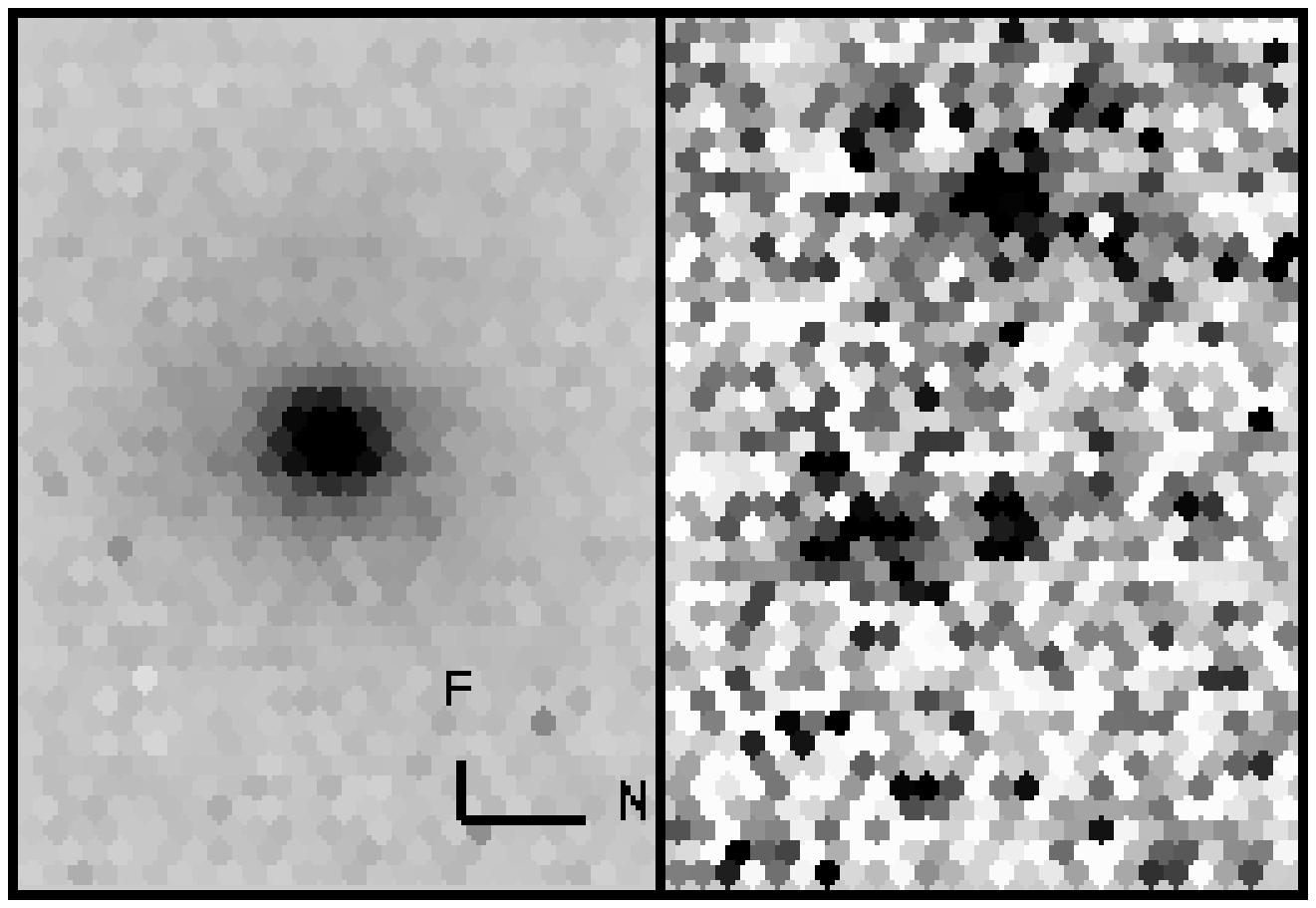}\hspace*{12pt}\includegraphics{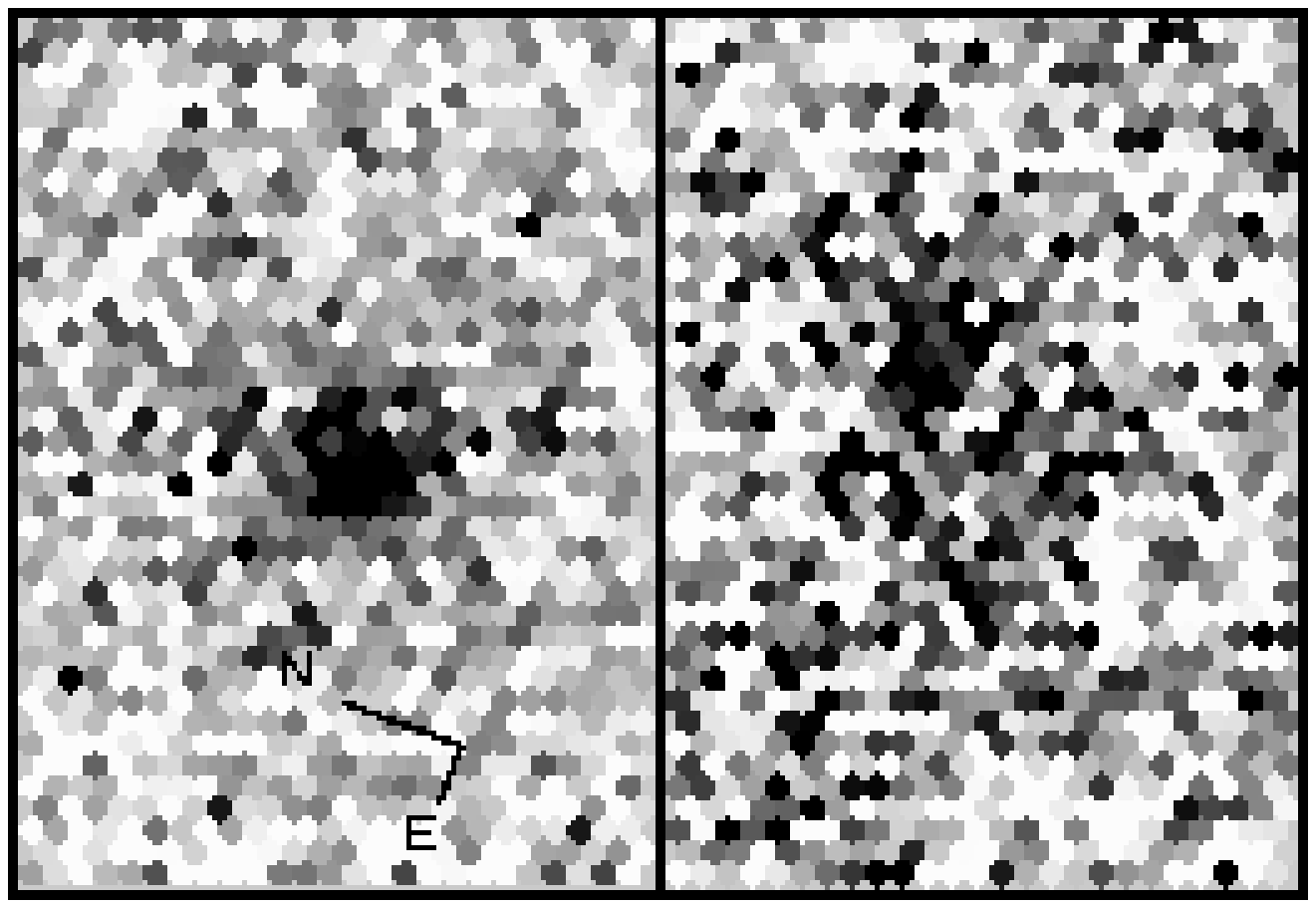}}}
\vspace*{5pt}
\centerline{\scalebox{0.38}{\includegraphics{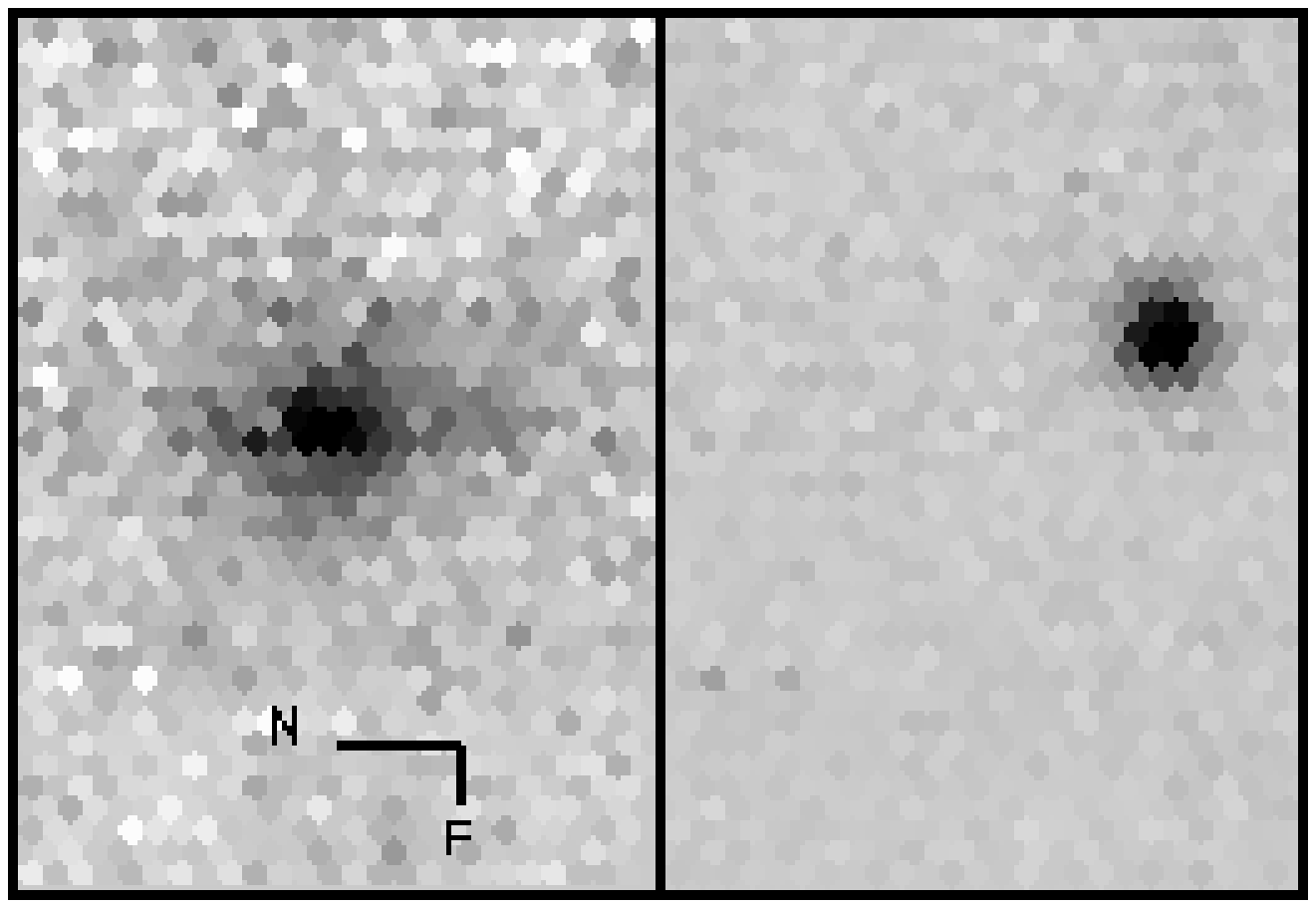}\hspace*{12pt}\includegraphics{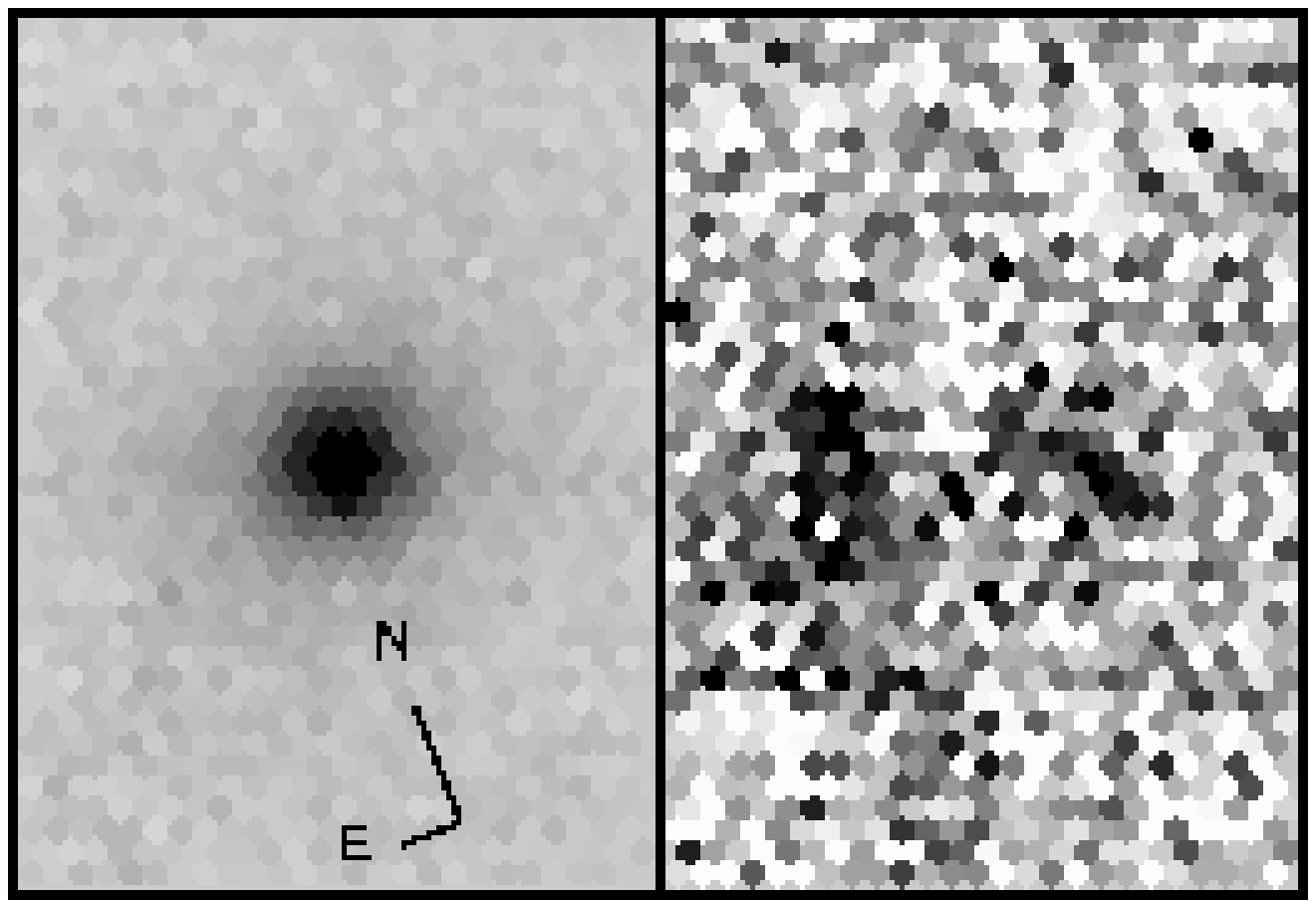}\hspace*{12pt}\includegraphics{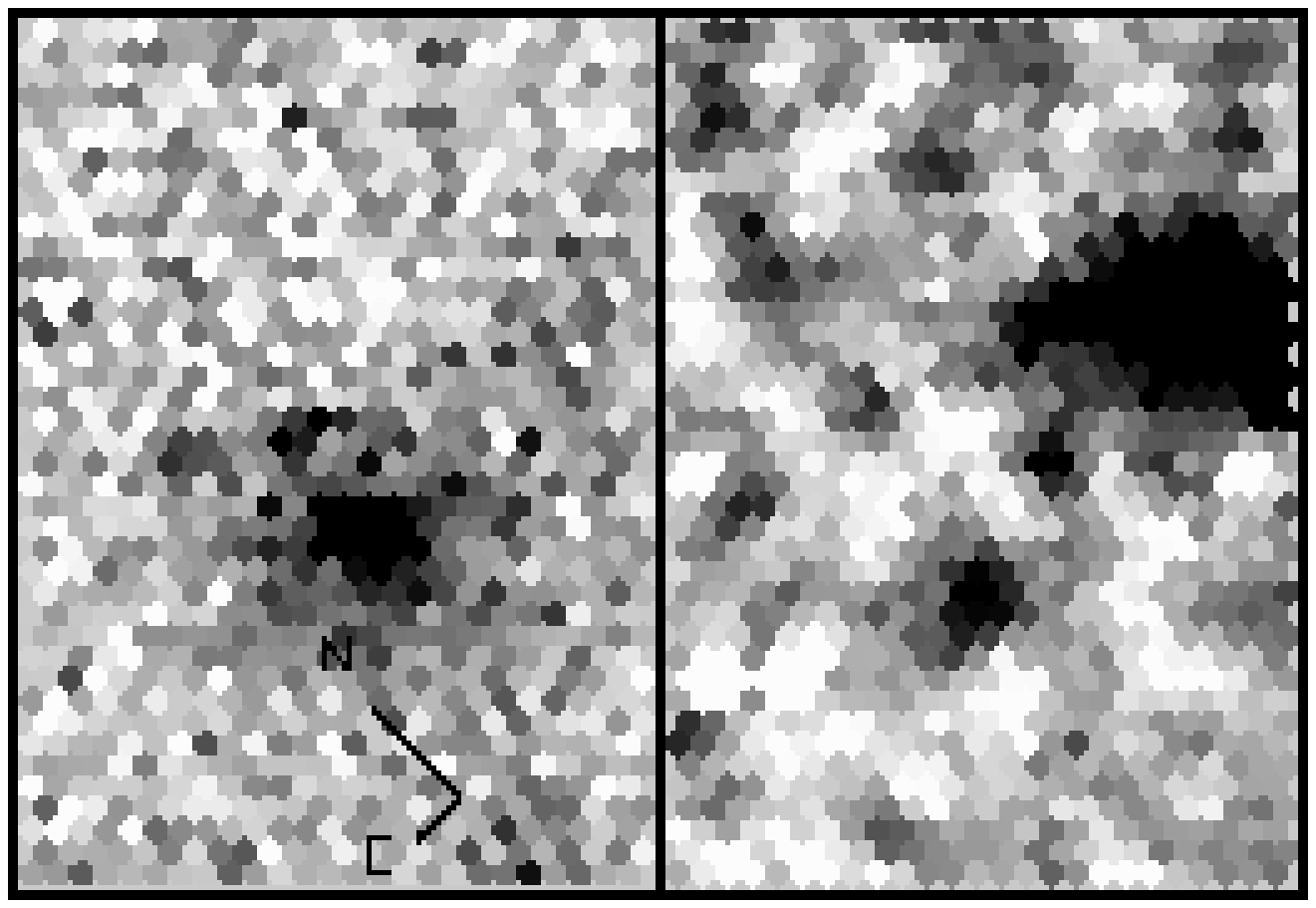}}}
\vspace*{5pt}
\centerline{\scalebox{0.38}{\includegraphics{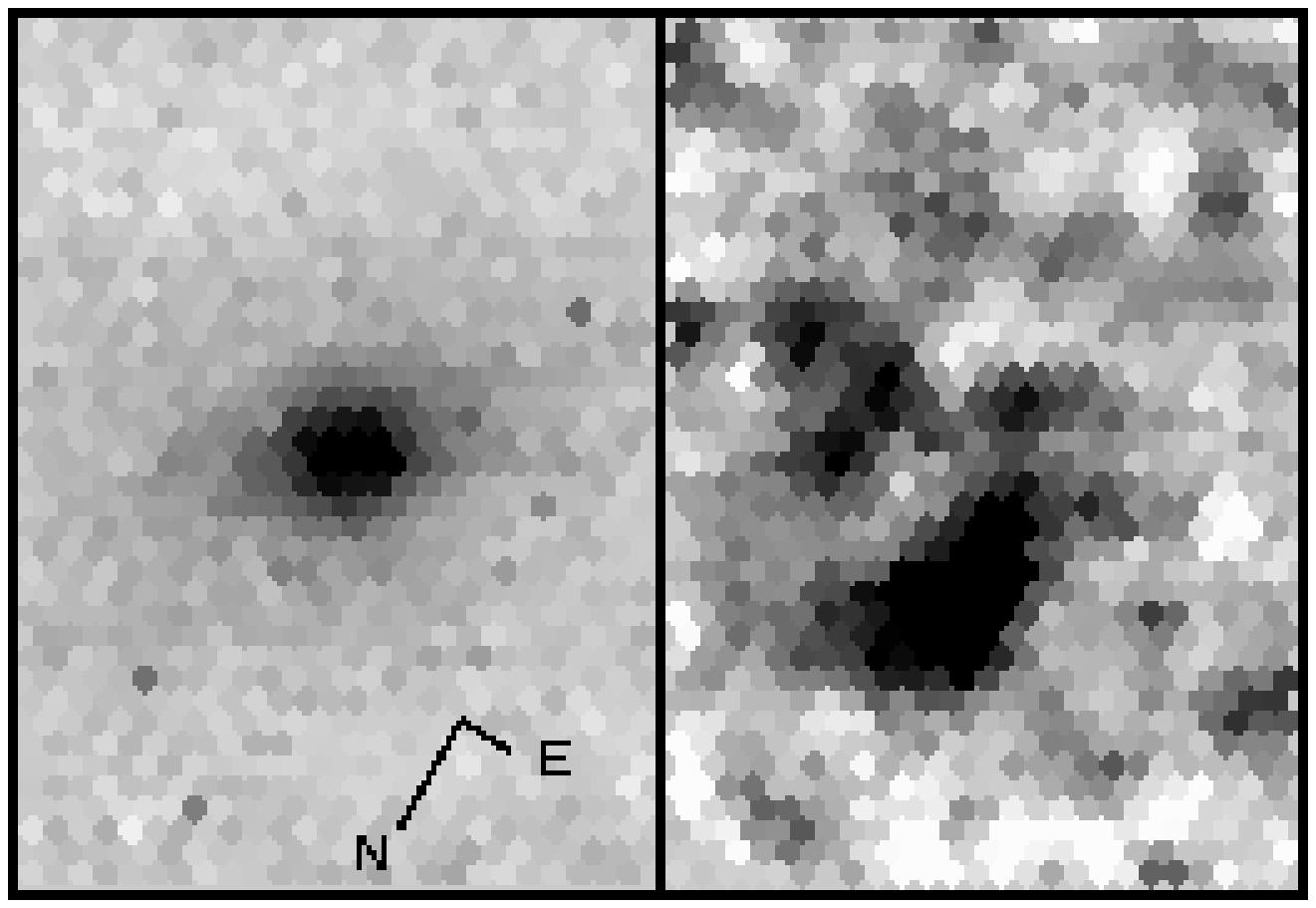}\hspace*{12pt}\includegraphics{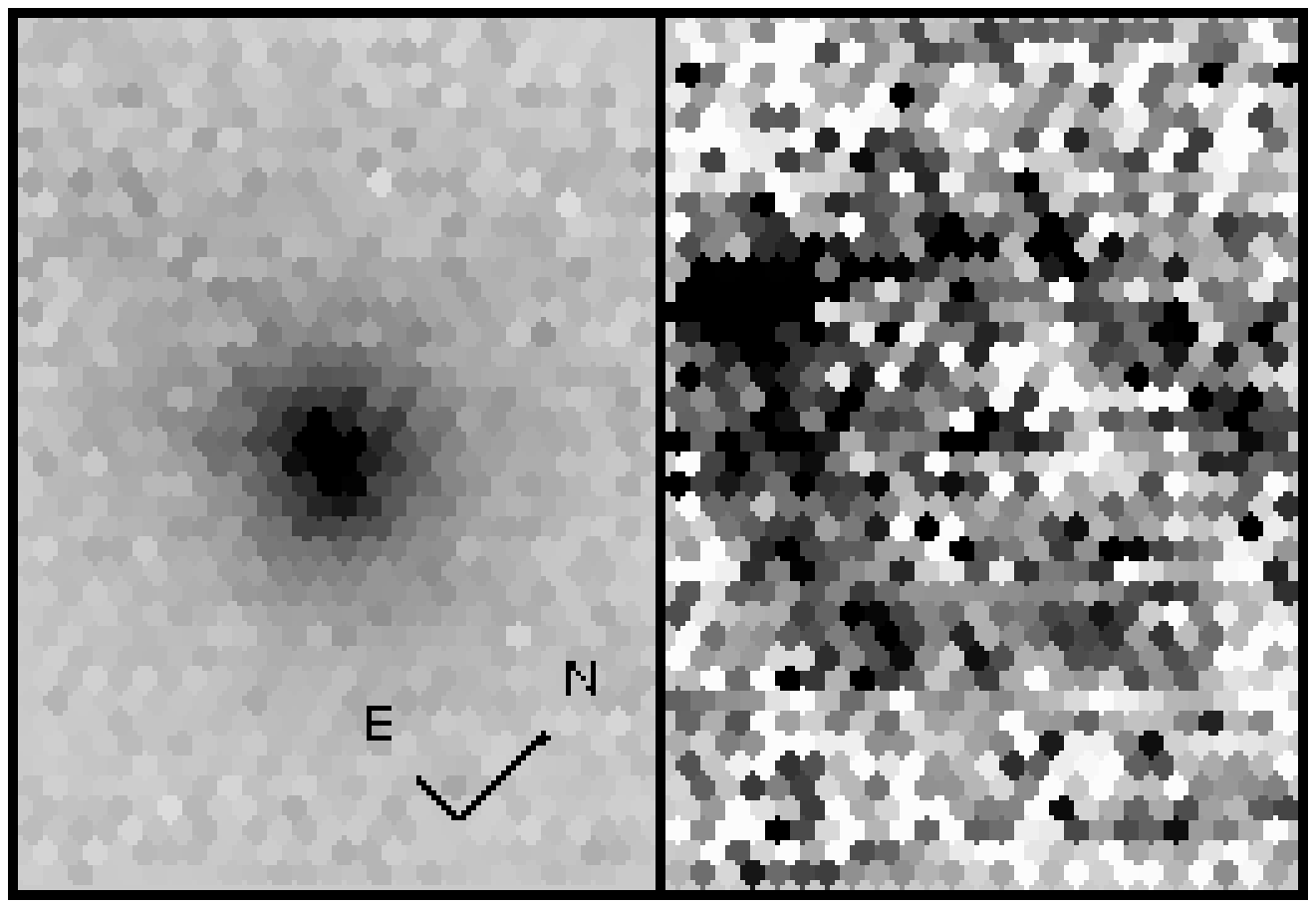}\hspace*{12pt}\includegraphics{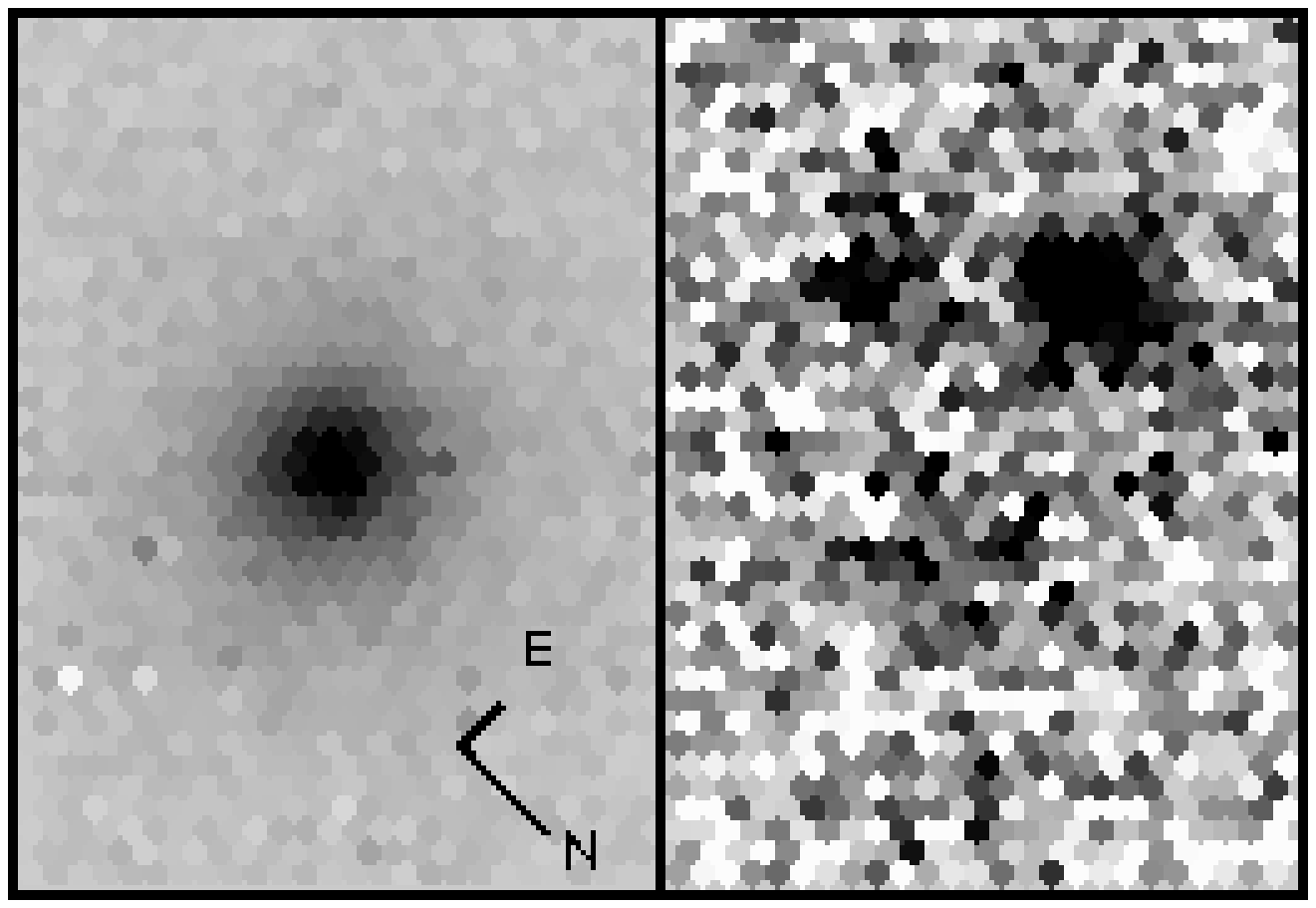}}}
\caption{\label{otherifu}
Systems with confirmed emission-line flux that do not admit
successful IFU gravitational-lens modeling.
Continuum (left) and emission-line (right) panels for each system
are as described in Figure~\ref{lensmodels}.
In reading order from upper left, the individual systems are:
J0928, J0956, J1128, J1155, J1259, J1416, J1521, J1630, J1702.
For J1416 and J1521, emission-line images have been smoothed
with a 7-fiber hexagonal kernel.  Discontinuity in continuum
image of J1155 is due to limited overlap of spectra from
the two GMOS-N IFU pseudo-slits degrading the object and calibration
data in the relevant wavelength range over one half
of the spatial field of view.}
\end{figure}

The linearly fitted
coefficients of the final parameterized basis functions are
our best decomposition of the spectrum into
background (emission-line) and foreground (continuum)
components.  Using these components, we can form emission-line
and continuum images in the
focal plane of the telescope using the known IFU field mapping.
Note that the modeling process yields optimal signal-to-noise
ratio for these images, better than simple flux-averaging
within defined continuum and emission-line wavelength aperture.
Figures \ref{lensmodels},~\ref{otherifu}, and~\ref{ifu1029} show reconstructed
narrow-band images of those systems for which the original SDSS
emission-line detection was confirmed by IFU observation, with
related information given in Table~\ref{datatab}.  IFU observations
of several additional candidate systems described in Table~\ref{nodetect}
failed to confirm the original SDSS line detection, and the associated
data are not presented in this paper.

\begin{table}
\caption{\label{datatab}
Observation and data display details for double-redshift systems with
IFU data presented in this paper.  Observation modes are described
in Table~\ref{modetab}.  ``IFU line'' column specifies the particular
background emission line used to generate the narrow-band images and models shown
in Figures \ref{lensmodels},~\ref{otherifu}, and~\ref{ifu1029} (with observed wavelengths
given by the background redshifts in Table~\ref{datatab}).  ``Figure levels''
columns give the limits of the gray-scaling used in
Figures \ref{lensmodels},~\ref{otherifu}, and~\ref{ifu1029}.
Units are 10$^{-17}$\,erg~cm$^{-2}$\,s$^{-1}$\,\AA$^{-1}$\,lenslet$^{-1}$
for continuum images and
10$^{-17}$\,erg~cm$^{-2}$\,s$^{-1}$\,lenslet$^{-1}$
for emission-line images.  ``Lens model?'' column specifies whether or
not a singular isothermal ellipsoid lens model was successfully fit to the
narrow-band emission-line data.}
\begin{indented}
\item[]\begin{tabular}{@{}lllllll}
\br
~    & ~    & ~    & ~    & \centre{2}{Figure levels} & ~ \\ \ns
~    & Obs. & Exp. & IFU  &        \crule{2}          & Lens \\
Name & mode & time & line &        Cont. & Line       & model? \\
\mr
  SDSSJ0037$-$0942 & IMACS & 2$\times$1200s & [O\textsc{iii}]~5007 & 0.30 & 3.00 & Yes \\
  SDSSJ0044$+$0113 & IMACS & 2$\times$900s & H$\alpha$ & 0.50 & 6.00 & Yes \\
  SDSSJ0737$+$3216 & GMOS-$i$ & 3$\times$900s & [O\textsc{iii}]~5007 & 0.10 & 1.70 & Yes \\
  SDSSJ0928$+$4400 & GMOS-$i$ & 3$\times$900s & [O\textsc{iii}]~5007 & 0.15 & 0.50 & No \\
  SDSSJ0956$+$5100 & GMOS-$i$ & 3$\times$900s & H$\beta$ & 0.20 & 0.20 & No \\
  SDSSJ1029$+$6115 & GMOS-$i$ & 3$\times$900s & H$\alpha$ & 0.20 & 0.15$^{\ast}$ & Yes$^{\ast}$ \\
  SDSSJ1128$+$5835 & GMOS-$r$ & 3$\times$900s & [O\textsc{ii}]~3727 & 0.12 & 1.00 & No \\
  SDSSJ1155$+$6237 & GMOS-$i$ & 2$\times$900s & [O\textsc{iii}]~5007 & 0.12 & 4.00 & No \\
  SDSSJ1259$+$6134 & GMOS-$i$ & 3$\times$900s & [O\textsc{iii}]~5007 & 0.25 & 0.25 & No \\
  SDSSJ1402$+$6321 & GMOS-$i$ & 3$\times$900s & [O\textsc{iii}]~5007 & 0.25 & 0.40 & Yes \\
  SDSSJ1416$+$5136 & GMOS-$r$ & 3$\times$900s & [O\textsc{ii}]~3727 & 0.12 & 0.60 & No \\
  SDSSJ1521$+$5805 & GMOS-$i$ & 3$\times$900s & [O\textsc{iii}]~5007 & 0.20 & 0.24 & No \\
  SDSSJ1630$+$4520 & GMOS-$r$ & 3$\times$900s & [O\textsc{ii}]~3727 & 0.15 & 0.30 & No \\
  SDSSJ1702$+$3320 & GMOS-$i$ & 3$\times$900s & [O\textsc{iii}]~5007 & 0.30 & 0.30 & No \\
  SDSSJ2238$-$0754 & IMACS & 3$\times$1200s & [O\textsc{iii}]~5007 & 0.30 & 0.80 & Yes \\
  SDSSJ2302$-$0840 & IMACS & 3$\times$900s & H$\alpha$ & 0.60 & 3.00 & Yes \\
  SDSSJ2321$-$0939 & IMACS & 2$\times$1500s & H$\beta$ & 0.80 & 0.40 & Yes \\
\br
\end{tabular}
\item[]$^{\ast}$ Emission-line images of SDSSJ1029$+$6115 shown in Figure~\ref{ifu1029}
are in units of 10$^{-17}$\,erg~cm$^{-2}$\,s$^{-1}$\,\AA$^{-1}$\,lenslet$^{-1}$.
See Section~\ref{threed} for full description of J1029 lens modeling.
\end{indented}
\end{table}

\begin{table}
\caption{\label{nodetect}
Lens candidate systems for which IFU observations
did not confirm the original SDSS emission-line detection.
All systems were observed with 3$\times$900s on GMOS-N.
Further information on these candidates is published in
\citet{bolton_speclens}.}
\begin{indented}
\item[]\begin{tabular}{@{}lllllll}
\br
Name & emission line & comment \\
\mr
SDSSJ1151$+$6455 & [O\textsc{ii}]~3727 & non-detection \\
SDSSJ1310$+$6211 & H$\beta$ & non-detect. [O\textsc{iii}]~5007 in atmos.\ absorp. \\
SDSSJ1550$+$5217 & [O\textsc{ii}]~3727 & non-detect. \\
\br
\end{tabular}
\end{indented}
\end{table}

\section{Strong lens morphology and modeling}
\label{modelsec}

The emission-line images of the
systems shown in Figure~\ref{lensmodels}, along with a number
of systems in Figure~\ref{otherifu} (specifically,
J0956, J1259, J1416, J1521, J1630, and J1702), show multiple features
that suggest strong gravitational lensing.
\citet{slacs1} show how many of these same
narrow-band IFU emission-line images confirm the lensing
interpretation of \textsl{HST} imaging data by showing
the spatial coincidence of the detected high-redshift
emission-line features with the putative lensed images
seen with \textsl{HST}.
Confirmation of a lens candidate as a gravitational lens
solely through IFU observation,
however, relies on the successful fitting of a lensing
mass model to the putative strongly lensed images.  Here we describe the
IFU lens modeling strategy that we apply to all possible
lens images, with successful results for the 7 systems shown
in Figure~\ref{lensmodels}.

Where the signal-to-noise ratio and spatial resolution allow,
we analyze all candidate lens systems uniformly by fitting
a singular isothermal ellipsoid mass model
(SIE: \citealt{kassiola_kovner, kormann_sie})
to the observed
emission-line image configurations.  The SIE model is
well motivated, simple, analytic,
and able to reproduce all qualitative
features of galaxy-scale strong lenses.  (A related model,
the singular isothermal sphere with external shear, also
offers these advantages.)
The convergence (scaled surface density)
of the SIE model is given by
\begin{equation}
\kappa_{\mathrm{SIE}} = {1 \over 2} {b \over {r_q}} ~~,
\end{equation}
where
\begin{equation}
r_q = \sqrt{q x^2 + y^2 / q}
\end{equation}
and $x$ and $y$ are angular coordinates
aligned along the major and minor axes of
the iso-density contours.
The dependence of the model on $x$ and $y$
only through $r_q$ means that all iso-density
contours are aligned, similar (not confocal), concentric ellipses.
The parameter $b$ expresses the strength of the
lens and $q$ gives the minor-to-major axis ratio of the
iso-density contours (hence $0 < q \le 1$ by
convention).  When $q = 1$, the SIE reduces to the
singular isothermal sphere (SIS) and $b$ is equal
to the angular radius of ring images of on-axis sources.
This ``Einstein radius'' is in turn
related to the velocity dispersion $\sigma$ of the lensing
distribution through
\begin{equation}
\label{bsigma}
b_{\mathrm{SIS}} = 4 \pi {{\sigma^2}  \over {c^2}}
{{D_{\mathrm{LS}}} \over {D_{\mathrm{S}}}} ~~.
\end{equation}
($D_{\mathrm{LS}}$ and $D_{\mathrm{S}}$ are angular-diameter
distances from lens to source and observer to source.)
For purposes of comparison between models with
differing $q$ values,
we adopt the same intermediate-axis normalization
as \citet{kormann_sie}, whereby the mass interior to a given
iso-density contour at fixed $b$
is constant with changing $q$.
A great advantage of the SIE is that its projected
gravitational potential can be expressed
analytically.  For our purposes we need only
the derivatives of this potential, which we use in the
simple form given by \citet{kk_spiral}.

We perform all lens modeling with our own IDL routines.
As is always the case
when fitting gravitational lens models, we must take
care not to fit for more parameters than are constrained
by the data.  Fortunately the SIE is generally free from
fundamental degeneracies among its parameters with regard to
the constraints furnished by real lenses.
In this paper
we are concerned with spatially extended
lensed star-forming galaxies which
will generally be of sufficient physical size to average over
any microlensing effects, and thus we use image surface
brightness rather than image positions
to constrain our lens models.
We begin the model-fitting procedure for each
individual system with a best-guess choice
for the parameters of the lens model and source galaxy
that gives a reasonable qualitative approximation to the
observed emission-line image morphology.
For systems that show spatially resolved emission-line
brightness distributions (most cases), we fit for lens and
source parameters by generating an unlensed source-galaxy
image (either Gaussian or S\'{e}rsic, with ellipticity
if necessary) and viewing it through the potential of
the parameterized lens model.  This image is then smeared
by a Gaussian PSF model (whose width is fit as another
free parameter), integrated over the hexagonal
IFU lenslets, and used to calculate $\chi^2$ directly relative
to the narrow-band IFU data.  The model parameters are
optimized non-linearly to minimize the $\chi^2$ statistic.
We expect appreciable degeneracies between best-fit
model parameters (i.e.\ between the PSF width and the intrinsic
source size), but the parameter of greatest interest to
us---the model's Einstein radius $b$---is largely
orthogonal to the others.
For some lenses, we constrain the center of the lensing
potential to be coincident with the center of the lensing galaxy,
which we determine by fitting a S\'{e}rsic model to the
IFU continuum image.
For one lens consisting of unresolved images (J0037), we fit the
images with hexagonally-sampled Gaussians (constrained
to have the same width as one another).
The image positions and fluxes
from these fits are then used to constrain the lens model,
with $\chi^2$ computed from the image position and flux residuals.
Table~\ref{modelpars} gives the measured angular Einstein radii that
result from our lens model fitting procedure.  Also shown are the
Einstein radii measured from a similar analysis of \textsl{HST}-ACS
imaging of these same systems (\citealt{slacs5};
see also \citealt{slacs3})

\begin{table}
\caption{\label{modelpars}
Lens strength $b$ (SIE Einstein radius) as measured by IFU lens modeling.
Quoted errors on the Einstein radii are
square-root-diagonal entries from the covariance matrix of the
nonlinear fit.
Also shown are \textsl{HST}-ACS $b$-values from \citet{slacs5},
where available.  IFU Einstein radii are converted to velocity dispersions
$\sigma_{\mathrm{SIE}}$ using Equation~\ref{bsigma}, which can be compared
with the stellar velocity dispersion $\sigma_{\star,\mathrm{SDSS}}$
measured from SDSS spectroscopy.}
\begin{indented}
\item[]\begin{tabular}{@{}lllcl}
\br
~ & $b_{\mathrm{IFU}}$ & $b_{\mathrm{HST}}$ & $\sigma_{\mathrm{SIE}}$ & $\sigma_{\star,\mathrm{SDSS}}$ \\
System & $(\arcsec)$ & $(\arcsec)$ & (km s$^{-1}$) & (km s$^{-1}$) \\
\mr
SDSSJ0037$-$0942 & 1.49$\pm$0.01 & 1.53 & 282 & 279$\pm$10 \\
SDSSJ0044$+$0113 & 0.73$\pm$0.01 & 0.78 & 259 & 266$\pm$13 \\
SDSSJ0737$+$3216 & 1.02$\pm$0.01 & 1.00 & 295 & 338$\pm$16 \\
SDSSJ1029$+$6115 & 0.60$\pm$0.01 & N/A  & 241 & 228$\pm$14 \\
SDSSJ1402$+$6321 & 1.41$\pm$0.01 & 1.34 & 300 & 267$\pm$17 \\
SDSSJ2238$-$0754 & 1.21$\pm$0.02 & 1.26 & 232 & 198$\pm$11 \\
SDSSJ2302$-$0840 & 1.03$\pm$0.01 & N/A & 248 & 237$\pm$ 8 \\
SDSSJ2321$-$0939 & 1.55$\pm$0.03 & 1.60 & 255 & 249$\pm$ 8 \\
\br
\end{tabular}
\end{indented}
\end{table}

\section{Three-dimensional lens modeling}
\label{threed}

\begin{figure}
\centerline{\scalebox{1.0}{\includegraphics{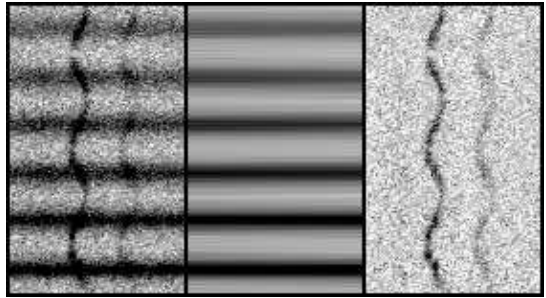}}}
\caption{\label{rotcurve}
A subsection of the extracted IFU spectra for J1029, running
from 8180\,\AA\ to 8270\,\AA\ (horizontal) and covering 150 lenslets
(vertical).  Left panel shows data, center panel shows fitted continuum
model, and right panel shows continuum-subtracted residual data.
Note the resolved velocity structure in the H$\alpha$ emission-line flux.
Second-most prominent line is [N\textsc{ii}].
Gray-scale is linear from -0.05 to 0.2
$\times 10^{-17}$\,erg~cm$^{-2}$\,s$^{-1}$\,\AA$^{-1}$\,lenslet$^{-1}$.}
\end{figure}

\begin{figure}[t]
\centerline{\scalebox{0.38}{\includegraphics{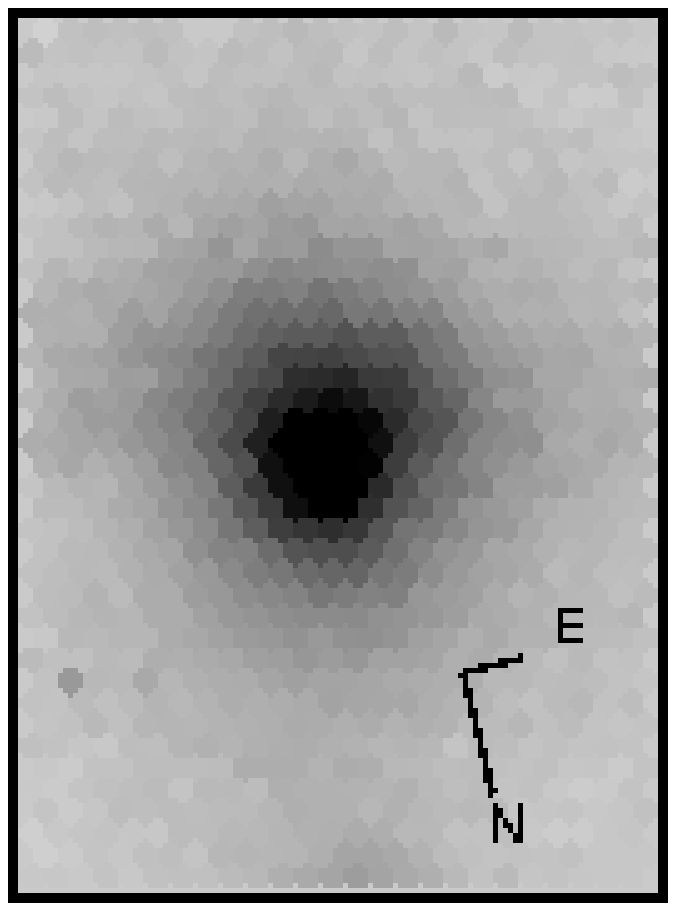}\hspace*{12pt}\includegraphics{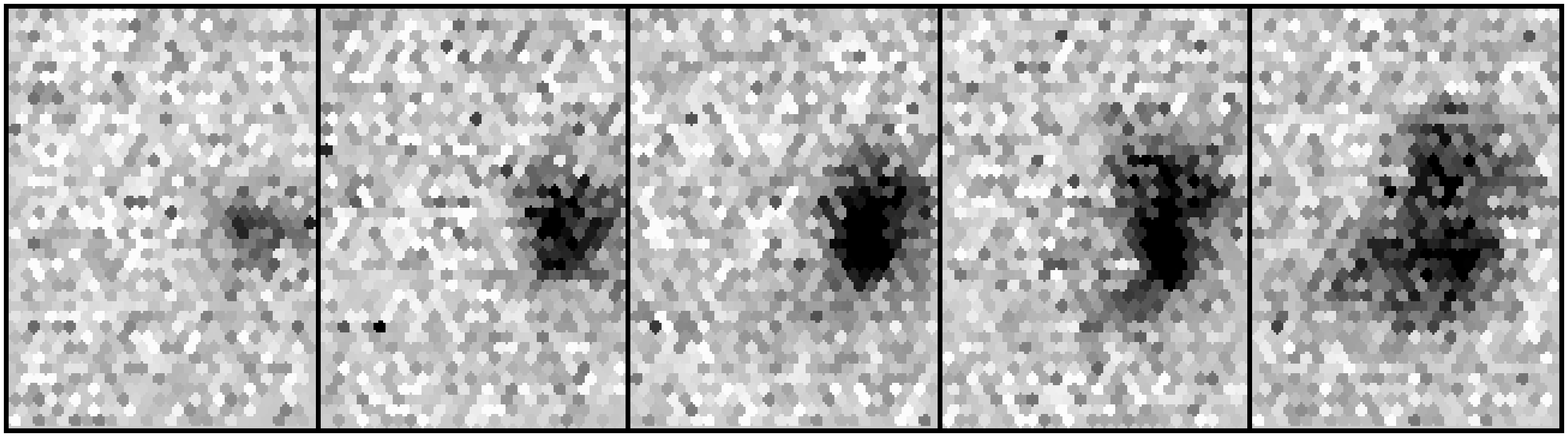}}}
\vspace*{3pt}
\centerline{\scalebox{0.38}{\includegraphics{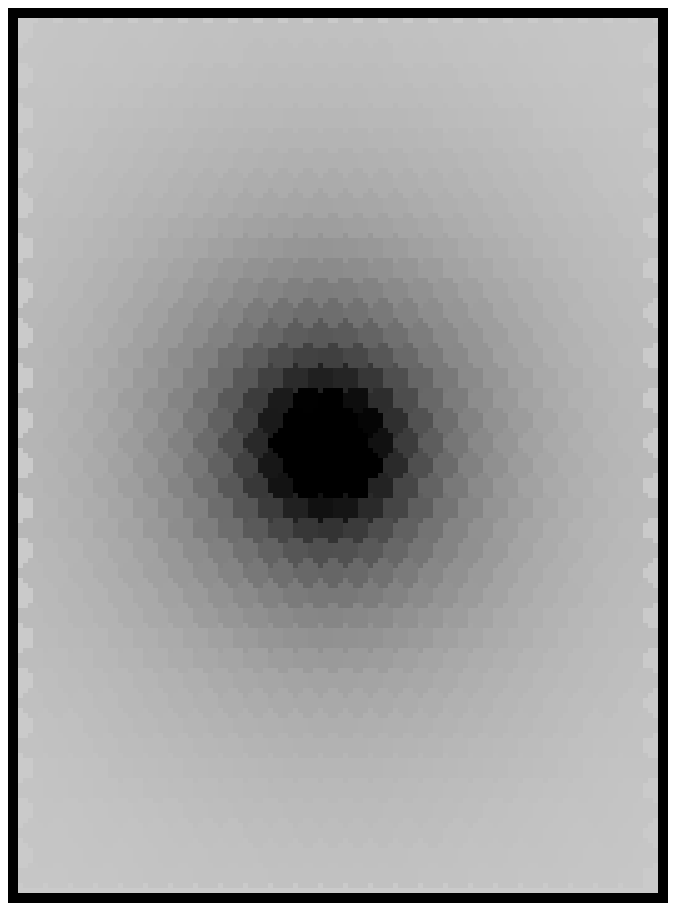}\hspace*{12pt}\includegraphics{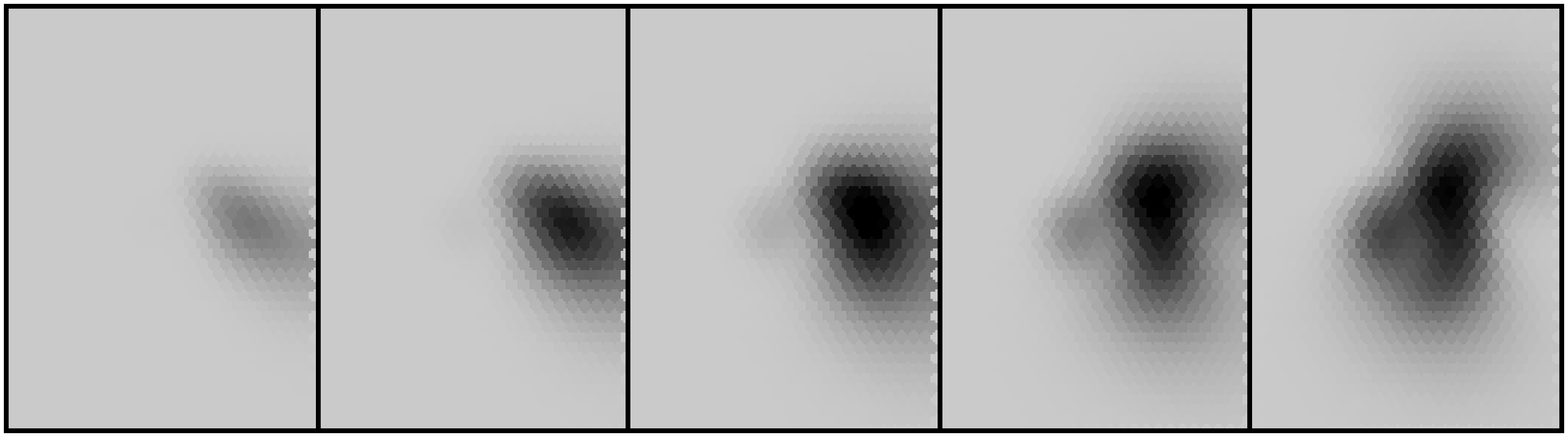}}}
\vspace*{6pt}
\centerline{\scalebox{0.38}{\includegraphics{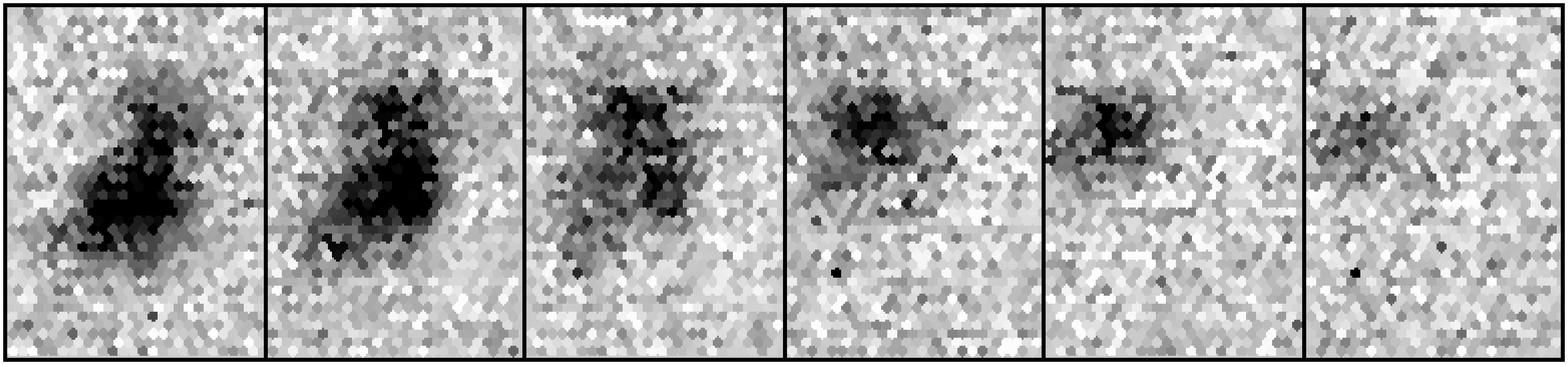}}}
\vspace*{3pt}
\centerline{\scalebox{0.38}{\includegraphics{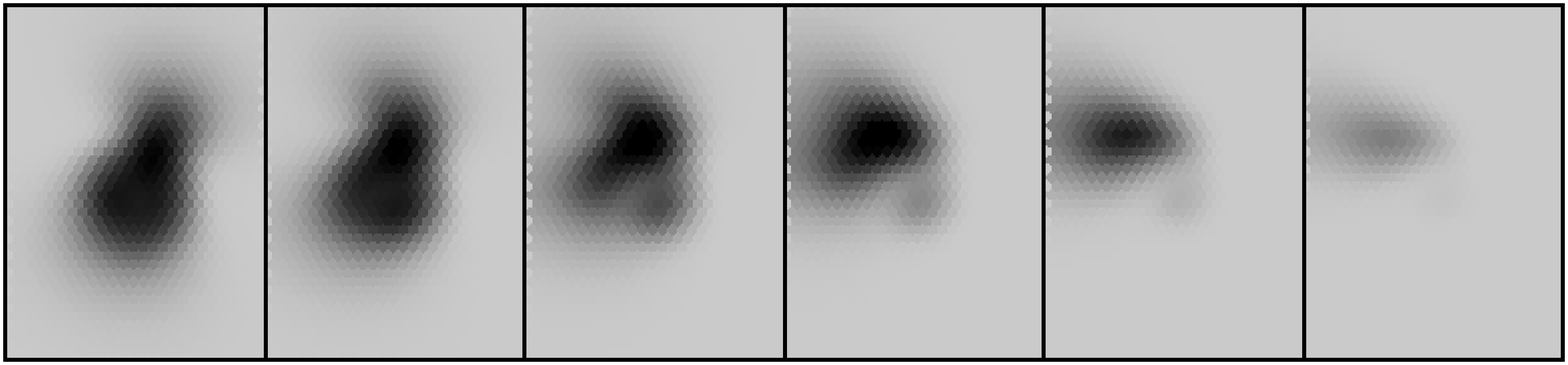}}}
\caption{\label{ifu1029}
Narrow-band IFU imaging of SDSSJ1029$+$6115 centered on the redshifted H$\alpha$
line of the background galaxy.  Leftmost panels in top two rows show continuum
data and Gaussian-convolved S\'{e}rsic-model fitted to those data, respectively.
Remaining panels show 1-\AA\ slices through the continuum-subtracted data cube
from 8210\,\AA\ to 8220\,\AA, with data above and lens model (described in
Section~\ref{threed}) below.  The model provides a reasonable qualitative
description of the data, though shortcomings are apparent.}
\end{figure}

With one exception, the internal velocity gradients in the
lensed emission line galaxies observed by this program are unresolved
spectroscopically.  The exception is J1029, for which the rotation
curve of the lensed galaxy is clearly visible.  Figure~\ref{rotcurve}
shows a section of the extracted IFU spectrum array for this
system.  The appearance of resolved velocity structure in the lensed
source affords a unique opportunity for three-dimensional strong lens
modeling: two spatial dimensions plus one velocity dimension.
Since multiple points in the lensed image plane coming from the same
point in the source plane not only have the same surface
brightness but also the same velocity, this technique
can in principle constrain the lensing model in more detail than
techniques using only velocity-unresolved data.  In practice, the
spatial resolution of our J1029 data is not sufficient to enable
a high-quality analysis.  Nevertheless, the wavelength-dependent
(i.e., velocity-dependent) lensed image structure,
seen in iso-wavelength slices through the continuum-subtracted data-cube
in the first and third rows of Figure~\ref{ifu1029},
reveals intriguing behavior that invites a plausible explanation
in terms of three-dimensional lens modeling.  We describe our
modeling technique here, which predicts the data-cube slices
shown in the second and fourth rows of Figure~\ref{ifu1029}.
Though the data and model disagree in detail, the qualitative
features of the data are reproduced by the model.

We adopt the same SIE model as above to describe the mass distribution
of the lensing galaxy.  We model the lensed background galaxy as an
infinitesimally thin, inclined exponential disk with a constant circular
orbital velocity
as a function of galactocentric radius and a single overall systemic average
velocity along the line of sight.  For any given set of parameters,
we compute the model image and velocity field of the background galaxy
as seen through the lens potential.  We then convolve the image and the
intensity-weighted velocity and squared-velocity images with the spatial
PSF and the spectral resolution, and assign to each IFU fiber in the model
data cube a Gaussian line profile with the appropriate line intensity,
central wavelength, and broadening.  The $\chi^2$ statistic is then
computed with respect to the continuum-subtracted data-cube, and the
parameters are optimized non-linearly.  The model shown in Figure~\ref{ifu1029}
represents the best result obtainable.  The mismatches in detail
between data and model are likely attributable to the inadequacy of the
parameterized model employed.  Unfortunately the limited spatial resolution of
the data prevent us from exploring more detailed models for the
source galaxy, which might in turn allow us to constrain more detailed models
for the mass distribution of the lens.  (In the simpler two-dimensional
modeling context above, we experienced similar difficulty with J1630.)

\section{Summary and conclusions}

We have shown that the high spatial resolution IFUs of the Gemini-N
and Magellan telescopes can be used successfully to confirm the
spatially extended nature of higher-redshift line emission detected along
the line of sight behind lower redshift elliptical galaxies.
The third (wavelength) dimension permits exceptionally accurate
decomposition of the reconstructed narrow-band images into continuum
and emission line components.  The emission-line images can then
be analyzed in a lensing framework virtually free from contamination
by the continuum of the foreground galaxy.  In a number of cases,
the IFU narrow-band emission-line images are sufficient
resolution and signal-to-noise ratio to
constrain gravitational-lens models that agree with models
from high-resolution \textsl{HST} imaging to within $0\farcs05$ RMS
(i.e., $1/4$ of a lenslet).
This result indicates that much---though certainly not all---of the
science that can be done with space-based imaging of these lenses
could perhaps be done with ground-based telescopes,
provided sufficient aperture, time, and image quality.
Space-based imaging enjoys a great advantage in the
interpretation and analysis of strong galaxy-galaxy
lenses with complex and irregular spatial morphology in the
lensed images.
This is the case for J1630, shown in Figure~\ref{otherifu},
which can be seen at \textsl{HST} resolution to consist of
five distinct and irregularly oriented knots of emission
in the source plane \citep{slacs1, slacs3}.
However, even in cases where space-based imaging is
clearly superior for lens modeling, IFU spectroscopy is of
great value as an independent confirmation of the lensing hypothesis.

We have also demonstrated the technique of three-dimensional strong
gravitational lens modeling, which is currently only feasible
using large ground-based telescopes such as Gemini and Magellan.
Although the demonstration data presented here are
of insufficient resolution for high-quality lens modeling, we consider
this application to hold great promise, particularly if it can be deployed
in combination with adaptive optics.

\ack

Based in part on data gathered with the 6.5 meter Magellan
Telescopes located at Las Campanas Observatory, Chile. \\

\noindent Based in part on observations obtained under program
GN-2004A-Q-5 at the Gemini
Observatory, which is operated by the
Association of Universities for Research in Astronomy,
Inc., under a cooperative agreement
with the NSF on behalf of the Gemini partnership:
the National Science Foundation (United
States), the Particle Physics and Astronomy Research
Council (United Kingdom), the
National Research Council (Canada), CONICYT (Chile),
the Australian Research Council
(Australia), CNPq (Brazil) and CONICET (Argentina).

\appendix

\section{Data calibration and analysis}

\label{ifu_red}
The format of the data delivered by the GMOS-N and IMACS IFUs to
their respective CCD-mosaic detectors is sufficiently
complicated to justify analysis with specially-developed software.
Furthermore, the scientific application of strong lens modeling
requires highly accurate wavelength calibration,
flat-fielding, and sky subtraction.
For these reasons, we have developed our own set of IFU
data-analysis software tools written in the IDL language.
Although there is an officially maintained and distributed package for the
reduction of GMOS IFU data under IRAF, there is no such package for IMACS.
Having developed our own IMACS routines, we preferred adapting them
to GMOS usage over using the official GMOS package.
We refer to our software
as ``{\tt kungifu}'' (kung eye eff you), and in this Appendix we
describe its function.  The {\tt kungifu} package can be
obtained for usage by other investigators by contacting
the authors, though the authors
are unable to offer user support beyond the distributed documentation.

\subsection{Bias subtraction and data formatting}
\label{biassub}
The IMACS detector consists of a mosaic of eight
$2048 \times 4096$-pixel CCDs.  The bias level
of raw frames varies between the CCDs, within each
CCD, and from one exposure to the next.  For any given
exposure, the bias is well characterized as the
sum of row-overscan and column-overscan terms.  Our
IMACS bias-subtraction routine first fits a smooth
b-spline model \citep{deboor_bsplines}
to the overscan region at the end of each row
for a specified breakpoint spacing.
This model is evaluated for all pixels in all rows of
the image and subtracted.
The IMACS bias level also has significant
column-dependence, and a similar bspline fit to the overscan
region at the end of each column is subtracted next.
The bias-subtracted and overscan-trimmed individual CCD images
are then stored to a single multi-extension FITS (MEF) file.
The detailed GMOS-N bias pattern is more stable that the IMACS
bias pattern, but cannot be estimated from the overscan alone.
Thus we perform GMOS-N bias subtraction using mean bias images,
and also subtract an overall average bias value for each
of the three GMOS-N CCDs from the overscan region to account for
slight variations.  In two-slit IFU mode, the central
CCD records spectra from both pseudo-slits.  We break
the central CCD into 2 logically separate images to
separate these two regions, and store them as separate
MEF extensions along with the first and
third CCDs, after bias
subtraction and overscan trimming.

\subsection{Flat-field modeling and tracing}
\label{mflat}
\begin{figure}[t]
\centerline{\scalebox{0.6}{\includegraphics{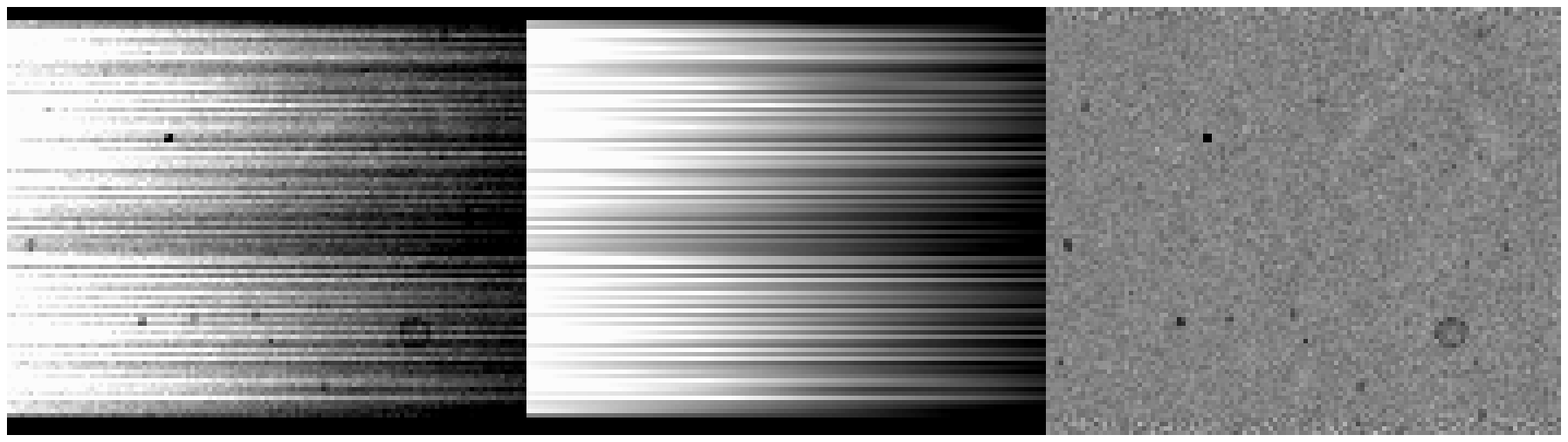}}}
\caption{\label{flat_fig}
Factorization of IFU spectroscopic flat-field
image via bspline modeling.  Left: small section of a
raw flat-field image for one 50-fiber IMACS-IFU block.
Spectra run horizontally; lighter color $=$ higher counts.
Center: model flat generated for same section by bspline fits to
each row, showing fiber response.  Right: ``pixel
flat'' for this section, given by ratio of raw flat
to model flat (left to center) and indicating relative pixel
response.}
\end{figure}

Relative flux calibrations of the IMACS and GMOS-N
IFUs are accomplished using dispersed IFU exposures of
uniform spatial illumination by (approximately)
flat-spectrum incandescent lamps.  Note that highly
uniform spatial illumination with facility calibration
equipment over such small IFU fields
of view is much more easily achieved than similarly
uniform illumination over
wide direct-imaging fields of view.
These raw spectroscopic flat images measure the product of
two non-uniform responses: the relative sensitivity
of individual CCD pixels, and
the fiber response image (including both individual fiber
profiles and relative throughput differences between fibers).
In an ideal spectrograph the illumination
pattern of the fibers would be fixed relative to the detector,
and these two effects would never need to be distinguished.
In actuality both GMOS-N and IMACS exhibit limited flexure
between successive exposures that
causes the IFU fiber spectra to shift their
position in CCD coordinates.  Thus we perform a factorization
of the pixel-response and fiber-response calibrations
by assuming an approximate scale separation between them.
Since the IFU fiber spectra for all data obtained for
this work run approximately along CCD rows,
horizontal cross sections through the IFU spectroscopic
flat frames follow the smooth variation of the
flat-lamp spectrum, modulated by gradual transitions
from one fiber to the next.  We generate smooth models
of the spectroscopic flat-field images by fitting
b-spline models to these
cross sections, with a breakpoint spacing chosen
ideally to be smaller than the typical scale of flat-lamp
features but greater than the scale of pixel-to-pixel
defects.  The resulting model-flat images are
an approximation to the fiber response, and the ratio of
raw- to model-flat images give the approximate
pixel response(``'pixel flat').
Figure~\ref{flat_fig} demonstrates this flat-field
factorization graphically.  We derive a master pixel flat
from the median-image of many individual pixel flats in a
given spectrograph configuration,
for application to all science frames in that
configuration.
We choose not to use imaging flats to calibrate the pixel
response, since this calibration can in general
be wavelength-dependent.  Using pixel flats derived from
spectroscopic flat-field frames ensures that the
correction is derived with illumination at the
appropriate wavelength.
We describe our use of the
model-flat images for the calibration of fiber response
in \S~\ref{extract} below.

We also use the model-flat images to determine the
location of individual fiber traces on the CCD\@.
For both the IMACS and GMOS-N IFUs, fibers along the
output pseudo-slit are grouped into blocks of 50,
with median inter-fiber spacings on the detector of
3.5
and 5.7
pixels respectively for
IMACS (in the $f/2$ or ``short'' camera mode)
and GMOS-N.
Fibers are approximately equally-spaced within the
blocks, and we use this fact to our advantage to
locate and trace all 50 fibers in a block at once.
For the purpose of locating block positions relative
to one another, our automatic tracing routine also
makes use of a table of inter-block
separations in units of the approximate local inter-fiber spacing
(which will be independent of pixel scale), which
is determined once for each IFU by a careful analysis
of a model-flat cross section.

\subsection{Scattered-light subtraction}

All IMACS and GMOS-N spectroscopic exposures exhibit
a non-negligible scattered-light background not directly
associated with the flux through the fibers, as evidenced
by non-zero count levels in the inter-block regions where
the flux from the fibers drops essentially to zero.
For all IFU frames used in this work---calibrations
and science exposures---we subtract a model of this background
after pixel-flat correction.  We estimate this scattered-light
image from the observed levels between the fiber blocks.
We use the flat-field-derived trace solution to define
bands in the inter-block regions running parallel
to the spectra, with a reasonable buffer to avoid the wings
of the fiber cross sections.  Each band is fit with a bspline
as a function of column, and this fit is subsequently evaluated
for each column.  We then interpolate this fit across the fiber
blocks by fitting a bspline in each column, taking the
band centers in that column as dependent variables
and the previous-fit evaluations in that column as independent
variables.  The b-spline breakpoint spacing in each
fit may be adjusted according to the signal-to-noise in
the scattered light levels and the degree of structure
that one wishes to model.

\subsection{Extraction}
\label{extract}

IMACS and GMOS-N
IFU observations distribute the photons from a few dozen
square arc-seconds of the sky over a few tens of millions
of CCD pixels.  Thus for all but the brightest objects,
low signal-to-noise is a primary concern.
This fact combined
with the well-behaved profile of the IFU
fibers on the CCD suggests optimal spectrum
extraction as a natural approach
\citep[e.g.][]{hewett_prism, horne_optimal}.
In an optimal extraction, the specific
flux of a spectrum at a given wavelength is
determined not from a simple sum over pixels
within a defined aperture, but rather from
the amplitude of a
maximum-likelihood fit of a model cross section
to the observed spectral cross section at that
wavelength.  This gives the maximum signal-to-noise
in the extracted spectrum, and is unbiased to the extent
that the model cross section matches the actual
cross section.  The most apparent obstacle to
optimal extraction of IMACS and GMOS-N IFU data
(aside from the sheer number of spectra)
is the significant overlap between neighboring spectra.
Fortunately the situation is less dire for fiber-fed
IFUs such as those of IMACS and GMOS-N than for
multi-object multi-fiber spectrographs,
since adjacent fibers on the detector are also
adjacent on the sky (an explicit design feature).
Since the $0\farcs 2$-diameter IFU lenslets
will critically sample all but the very best
ground-based seeing, the blending of
neighboring fiber spectra on the detector
leads to no significant loss of information
\citep{jas_gmos_ifu}.
Furthermore, modeling of the fiber profile is unnecessary,
since the model flats described in \ref{mflat}
provide us with a high signal-to-noise determination
of the spectrum cross-sectional shape for all fibers
and at all wavelengths.

Before using the model flats for
the extraction of spectra from the pixel-flat
corrected and scattered-light subtracted object
frames, we first normalize the model flats
(also scattered-light subtracted)
by dividing out a crude approximation to the flat-field
lamp spectrum as a function of wavelength.
This step is not crucial (particularly if one
eventually performs an absolute flux calibration),
but it prevents the flat-lamp spectrum from being imprinted on
the data before flux calibration.
We note that it is not important to use an exceedingly
accurate model of the lamp spectrum, but only to
divide all pixels with the same wavelength by the
same value.  Next we shift the model flats
perpendicular to the dispersion direction with
a flux-conserving damped-sinc kernel so as to maximize
the cross-correlation between the model-flat image
and the object-frame image to be extracted.  This shifting of the
model flat accounts for the slight (typically
of order 1 pixel or less) flexure that can occur between
the object frames and the flat frames taken immediately
following.

\begin{figure}[t]
\centerline{\scalebox{0.4}{\includegraphics{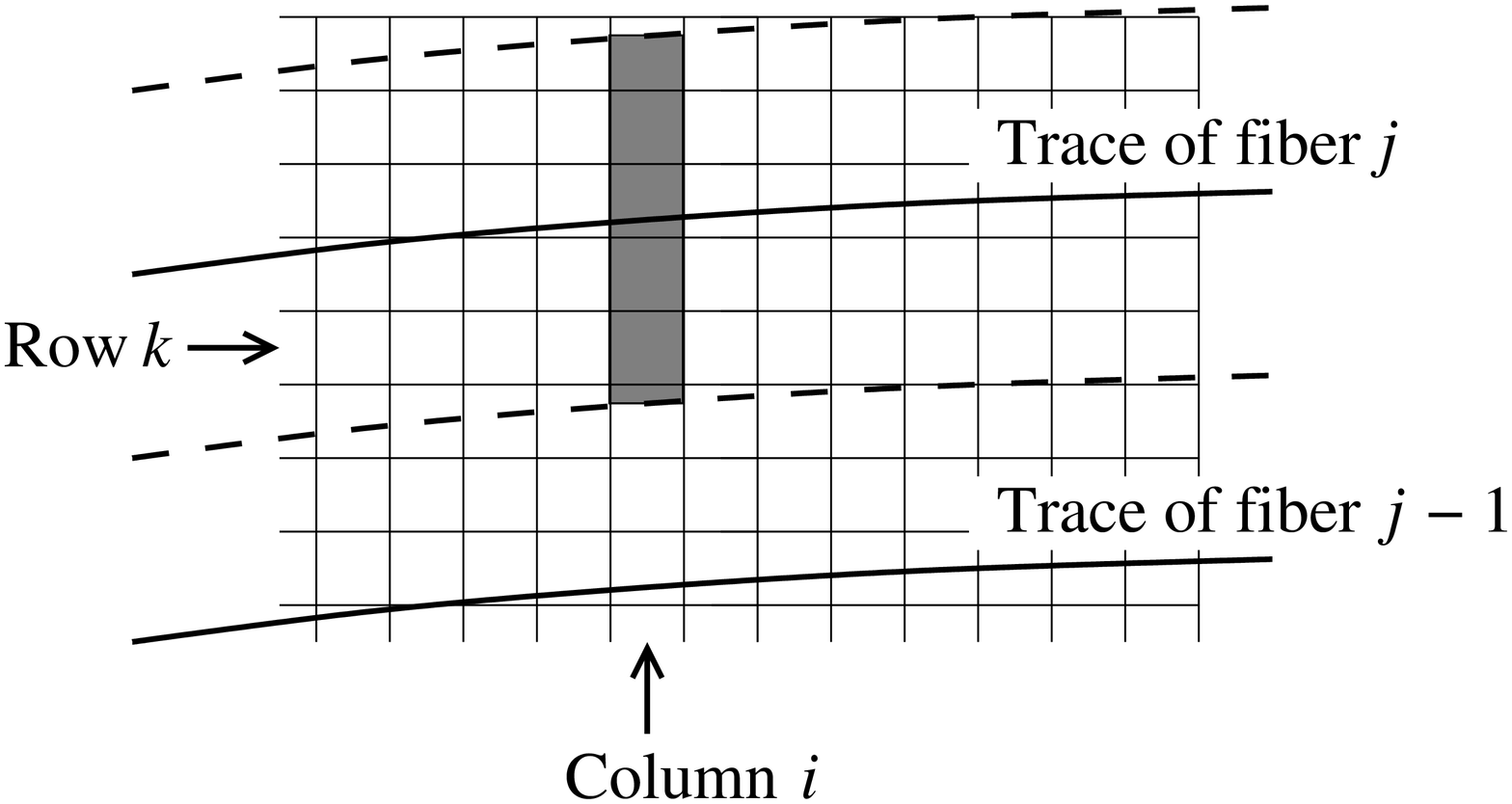}}}
\caption{\label{flat_fit}
Diagram of association of pixels with a
specific fiber $j$ in a specific CCD column $i$,
shown in gray.  Tilt of traces with respect to CCD
pixel grid is exaggerated.}
\end{figure}

Our approach to extraction
is described mathematically
as follows.  We define boundaries
between fiber spectra by lines exactly half-way
between the fiber traces, and in each CCD column $i$
we associate with fiber $j$
all pixels (i.e.\ rows)
$k$ falling between the boundaries on either side.
Pixels split by the boundary are associated fractionally
(see Figure~\ref{flat_fit}).
Let $w^{(i,j)}_k$ express this weighting:
$w^{(i,j)}_k = 1$ for rows $k$ wholly
within the boundaries for
fiber $j$ in column $i$, 0 for rows wholly outside, and
between 0 and 1 for rows fractionally included.
Let $d_{ik}$ be the data frame to be extracted,
$\sigma^2_{ik}$ be the statistical variance of $d_{ik}$,
and $f_{ik}$ be the aligned,
normalized model flat-field image of the fiber response.
The optimally extracted specific flux in
fiber $j$ at column $i$ (corresponding to a particular
wavelength by the dispersion solution for that fiber),
which we denote $I_{ij}$, will be given by the value that
minimizes
\begin{equation}
\chi^2 = \sum_k w^{(i,j)}_k (d_{ik} - I_{ij} f_{ik})^2 / \sigma^2_{ik} ~,
\end{equation}
which is
\begin{equation}
I_{ij} = \sum_k w^{(i,j)}_k f_{ik} d_{ik} / \sigma^2_{ik}
 \left/ \sum_k w^{(i,j)}_k f_{ik}^2 / \sigma^2_{ik} \right. ~.
\end{equation}
A simple adjustment of this expression suggests
a more succinct conceptual and operational approach:
\begin{equation}
I_{ij} = \sum_k (d_{ik} / f_{ik}) (w^{(i,j)}_k f_{ik}^2 / \sigma^2_{ik})
 \left/ \sum_k w^{(i,j)}_k f_{ik}^2 / \sigma^2_{ik} \right. ~.
\end{equation}
This instructs us to obtain the optimally extracted
specific flux by dividing the data image by the
model-flat image, then computing a weighted average
over the appropriate fiber/column window, with the
statistical weight given by the product of the
squared model-flat image and the inverse-variance image.
We implement the extraction algorithm in this manner.
The flat-fielding procedure thus yields calibrated
specific intensity measurements (flux per unit wavelength per unit area)
in a relative sense at any given wavelength across the
entire spatial field, though the absolute extracted
flux values are not meaningful unless and until an absolute
flux calibration is applied.

The optimal-extraction technique also provides
a natural means for rejecting cosmic-ray (CR) hits
in single exposures, because cosmic rays generally will
not have the same shape as the fiber cross section.
We flag pixels with highly statistically significant
positive deviations between the data frame and
the optimal-extraction model frame as CR pixels
and grow the resulting CR mask to laterally-adjacent
pixels, then repeat
the extraction with CR pixels given zero weight.

We store the extracted IFU object spectra for
each CCD in the IMACS and GMOS-N mosaics as a separate
image extension in a single MEF file for a single exposure.
These spectrum images have a horizontal dimension
equal to that of the CCD and a vertical dimension
equal to the number of fibers with spectra
falling on the CCD, and are not rectified in wavelength
but rather have a unique wavelength sampling for each
spectrum. We note here that
storing the data in this two-dimensional spectrum-image form
instead of in the ``data-cube'' form of two spatial dimensions
by one spectral dimension affords a distinct advantage even for
wavelength-rectified data (\ref{rect}) in
that it preserves the integrity of the detector frame.

\subsection{Relative and absolute wavelength calibration}
\label{wavecal}

We establish wavelength calibration using exposures
of He-Ne-Ar (for IMACS) and Cu-Ar (for GMOS-N) arc lamps.
We first process arc images by
subtracting the bias level,
dividing by a pixel flat, and subtracting
a scattered-light image model.
We then use the most closely-associated (in time)
model spectroscopic flat,
globally sinc-shifted perpendicular to the dispersion direction
if needed for alignment, to perform an optimal extraction
as described in \S~\ref{extract}.  The model flat image is not
normalized in wavelength since the wavelength calibration
is as yet unknown.
Within the resulting set of extracted arc spectra, we identify
as many individual bright lines as possible.
The centroids of these lines are found in each
spectrum through an iterative linearized
Gaussian centroiding algorithm that also measures
the width of each line in the dispersion direction.
The median line-spread width for each fiber is saved
for later use.

Due to global curvature of the pseudo-slit image on the
detector mosaic, each fiber spectrum has its own unique
dispersion solution, as would be the case for the individual
rows in a single long-slit spectrum.  Furthermore,
discrete offsets in the dispersion direction of
50-fiber blocks relative to one another, as well as offsets
of individual fiber spot positions from the mean within
their blocks, make a global 2D dispersion solution impractical.
Rather, we use the set of arc line positions in each fiber
to derive the individual dispersion solution in that fiber
{\em relative to a baseline defined by the average arc-line
position across all fibers}, which we refer to as the
``pixel-wavelength'' baseline.  We derive this solution in
the following iterative fashion.  First, for each arc line,
we compute the simple average centroid position (ccd column)
across all fibers, which we take to be the pixel-wavelength
of that line.  We then fit the position of all lines in each
fiber spectrum with a low-order (linear or quadratic) polynomial
function of this single pixel-wavelength baseline.  We then use
the average residuals of these fits across all fibers to correct the
adopted pixel-wavelength values for each line and re-fit. 
This process converges rapidly after a few iterations, and the
resulting polynomial solutions are accurate to within a few
hundredths of a pixel RMS difference between
arc-line centroids as measured and as predicted by the fit
(equivalent to a few hundredths of an Angstrom
RMS at the dispersions used in this work).
This process---which amounts to the determination, but not the
application, of a wavelength-rectifying solution---has several
advantages.  First, one need not know the identifications of
the arc lines used.  Second, one need not exclude
blended lines, since any bias introduced by blended lines into
the independent variable (pixel) will be exactly canceled
in the dependent variable (pixel-wavelength).  Finally, the
strategy reduces absolute wavelength calibration to a one-dimensional
problem.

Once the solution for pixel as a function of pixel-wavelength
has been determined for each individual fiber, we determine
the single solution for physical wavelength as a function of
pixel wavelength.
We first use the (inverse) pixel-wavelength
solution to determine the central pixel-wavelength of each pixel in
the extracted arc frame.  We then perform a single b-spline fit
to the extracted arc flux as a function of pixel wavelength across
all fibers.  From a signal-to-noise point of view we are effectively
stacking all the individual arc spectra,
but rather than re-binning/rectifying in (pixel-)wavelength,
we fit to the data in their native pixel sampling.
We identify peaks in the fitted b-spline model, and fit a
polynomial solution for the absolute wavelength calibration.
A further benefit of our decomposition of the wavelength calibration
problem into relative and absolute steps is that even arc lines detected
at SNR $\sim < 1$ in individual fiber spectra can be robustly
centroided in the global b-spline model and used in the fit.
Note that the separation into relative and absolute wavelength
calibration is entirely analogous to the usual separation of
flux calibration into the relative step of flat-fielding and
a subsequent step of absolute calibration using a flux standard.

When applying the wavelength calibration to the science data,
we also correct for slight
flexure between the object frames and the dispersion
solution from the calibrating arc frame
by fitting for a low-order two-dimensional polynomial transformation
as defined by the positions of known night-sky emission lines.

\subsection{Sky subtraction}
\label{skysub}

\begin{figure}[t]
\centerline{\scalebox{1.0}{\includegraphics{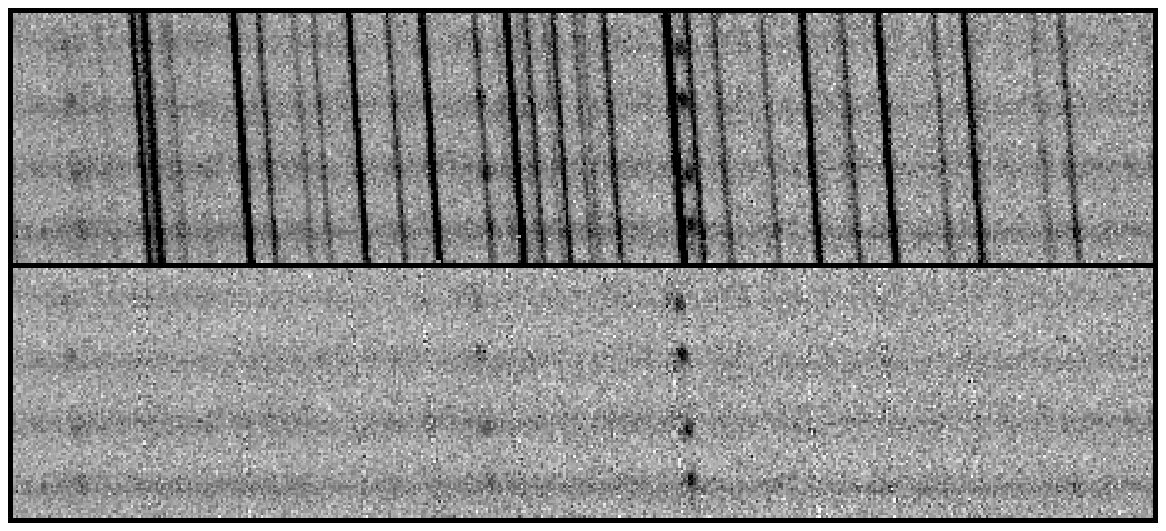}}}
\caption{\label{sky_sub}
GMOS-N $i$-band sky subtraction for a single 900s exposure:
before (above) and after (below).  A total of 100 fiber spectra
are displayed in each case, arranged vertically.  Wavelength
direction is horizontal, and covers approximately the same
range as is shown in Figure~\ref{sky_signif}.}
\end{figure}

\begin{figure}[t]
\centerline{\scalebox{0.65}{\includegraphics{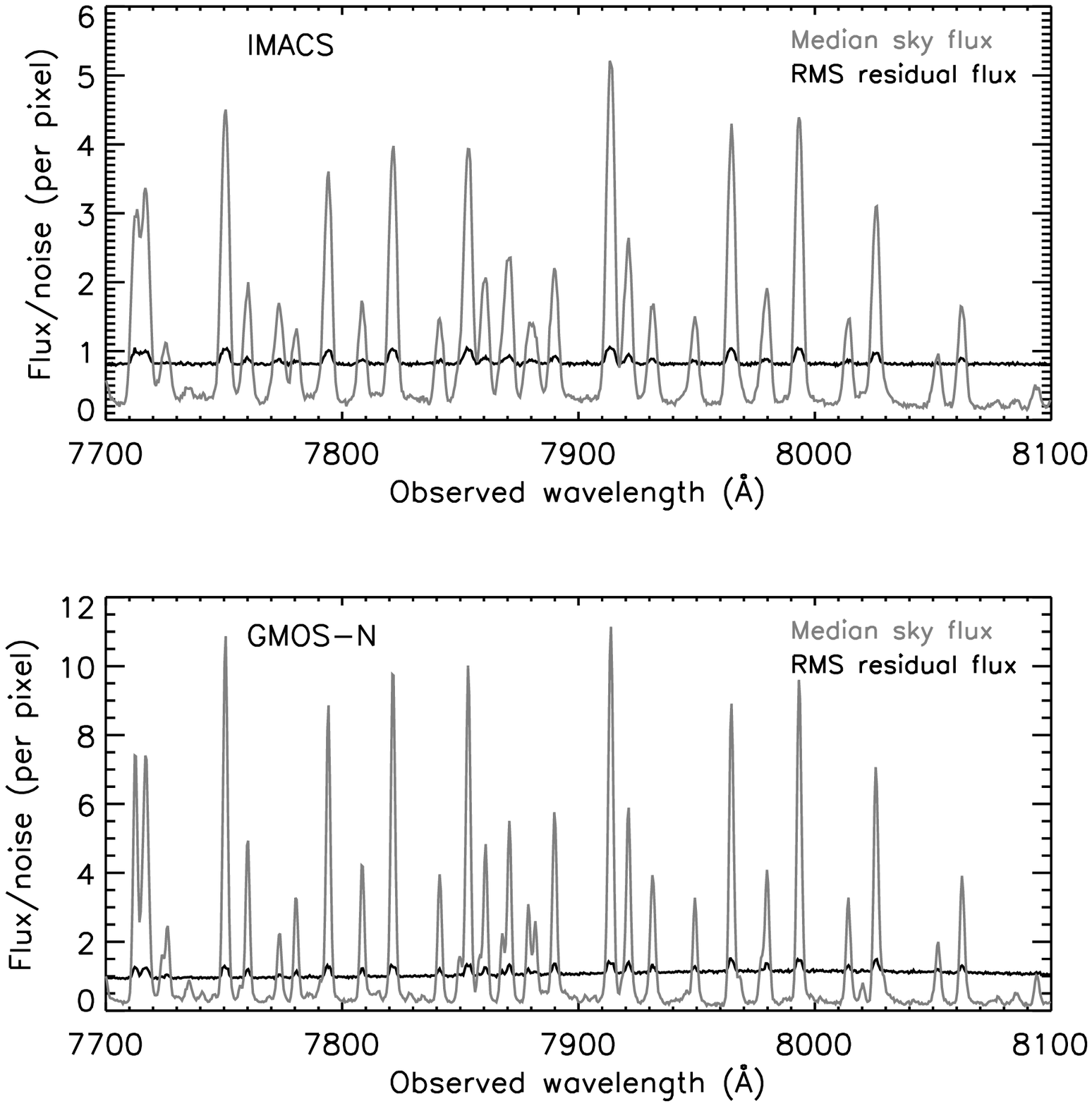}}}
\caption{\label{sky_signif}
Wavelength-dependent
IFU sky subtraction performance for IMACS (top) and GMOS (bottom)
in an $i$-band wavelength range containing strong OH
rotational transition lines.
Shown in gray is the median across fibers of the ratio
of sky counts to noise counts.  The RMS across fibers
of the ratio of sky-subtracted residual counts to noise counts
is shown in black.  When the noise is estimated correctly and
the sky is subtracted perfectly, this second ratio will be 1.
Pixel size is approximately 1.3\AA\ for IMACS and 0.9\AA\ for GMOS-N
(including a binning factor of 2).  Curves are computed for each
exposure separately using all sky fibers; curves shown are a
median across all individual exposures used in this paper.
No adjustment is made for differing exposure times, airmasses,
or moon phases between the individual exposures.}
\end{figure}

Both the GMOS-N and IMACS IFUs incorporate
dedicated fiber bundles for the observation of blank
sky, from which to estimate the foreground level to be subtracted
from the science data.
As discussed by \citet{kelson_skysub},
there is a distinct advantage in the estimation and
subtraction of the night-sky spectrum before performing
any rectification in wavelength.
The multiple fibers
of the IFU fields of view each have a slightly
different wavelength sampling on the detector, and hence the
discretely sampled line-spread function (LSF) observed for
night-sky emission lines depends upon the sub-pixel
location of the line's central wavelength.  The native binning of
the CCD, when considered for all blank-sky fibers together,
provides a finely-sampled observation of the night sky spectrum.
We thus fit a bspline model to this data as a function of the
central pixel-wavelength of each native spectral pixel, which we then
evaluate for all fibers (object and sky) and subtract as our
sky model.  The extracted one-dimensional LSF
of the IMACS and GMOS-N IFUs
exhibits some variation due to global
distortions and fiber-optic heterogeneity, which we
characterize via the LSF width
measured for each fiber from arc frames as described in
\ref{wavecal} above.  This LSF width is then treated
as a second independent variable in the bspline model,
fitted with linear or quadratic dependence.
Figure~\ref{sky_sub} shows the results of $i$-band sky subtraction
for a single 900-s GMOS-N frame, and Figure~\ref{sky_signif}
characterizes the typical statistical quality of sky subtraction
across the observational sample.

\subsection{Rectification and combination}
\label{rect}
To facilitate the combination of multiple exposures
(and to make analysis more straightforward), we rebin
our sky-subtracted IFU spectra onto a uniform
wavelength baseline.
We use a constant-wavelength binning across the spectrum, with
bin size slightly larger than the largest native pixel-width,
so that only nearest-neighbor correlations will be present
in the rebinned frames.  The wavelength-bin boundaries are
specified in heliocentric vacuum wavelengths, corrected
with a heliocentric velocity shift appropriate to each
observation, converted to air wavelengths, and
mapped into the extracted IFU frames
using the arc-frame dispersion solution, as adjusted
to match the observes night sky lines.
Multiple exposures are then combined
after re-binning, with further cosmic-ray rejection.
Finally, the rebinned data from individual CCDs are
combined onto a single mosaic image that maintains the
native orientation of the spectra
on the detector mosaic.  As a data product,
we prefer this mosaic
to a three-dimensional ``data cube'' (2 spatial plus 1 spectral
dimension) because it allows the reduced data to be displayed all
at once in the frame of the detector.  Data cubes may always
be constructed using the IFU field-mapping.
We do not rebin our data spatially, since all observations
were made with single undithered telescope pointings.

\subsection{Absolute flux calibration}

Flux calibration is currently not implemented as
an integral aspect of the {\tt kungifu} tasks,
and is not necessary for gravitational-lens modeling using
emission-line images at a single wavelength.
In order to judge the depth of our observations and the
scale of emission-line luminosities observed,
we implement a somewhat crude flux calibration for our gravitational-lens
candidate sample through a bootstrap connection to the flux-calibrated
SDSS spectra, which were obtained through a single
$3\arcsec$-diameter fiber aperture that can be synthesized from within
the IFU field of view.

%\bibliography{bolton_njp_07}
\References

\bibitem[{{Allington-Smith} {et~al.}(2002)}]{jas_gmos_ifu}
{Allington-Smith}, J., {et~al.} 2002, \pasp, 114, 892

\bibitem[{{Bigelow} \& {Dressler}(2003)}]{bigelow_imacs}
{Bigelow}, B.~C., \& {Dressler}, A.~M. 2003, in Instrument Design and
  Performance for Optical/Infrared Ground-based Telescop es. Edited by Iye,
  Masanori; Moorwood, Alan F. M. Proceedings of the SPIE, Volume 484 1, pp.
  1727-1738 (2003)., 1727--1738

\bibitem[{{Bolton} {et~al.}(2005){Bolton}, {Burles}, {Koopmans}, {Treu }, \&
  {Moustakas}}]{bolton_1402}
{Bolton}, A.~S., {Burles}, S., {Koopmans}, L.~V.~E., {Treu }, T., \&
  {Moustakas}, L.~A. 2005, \apjl, 624, L21

\bibitem[{{Bolton} {et~al.}(2006){Bolton}, {Burles}, {Koopmans}, {Treu }, \&
  {Moustakas}}]{slacs1}
---. 2006, \apj, 638, 703

\bibitem[{{Bolton} {et~al.}(2007){Bolton}, {Burles}, {Koopmans}, {Treu },
  {Moustakas}, \& {Gavazzi}}]{slacs5}
{Bolton}, A.~S., {Burles}, S., {Koopmans}, L.~V.~E., {Treu }, T., {Moustakas},
  L.~A., \& {Gavazzi}, R. 2007, \textit{in preparation}

\bibitem[{{Bolton} {et~al.}(2004){Bolton}, {Burles}, {Schlegel}, {Eisenstein},
  \& {Brinkmann}}]{bolton_speclens}
{Bolton}, A.~S., {Burles}, S., {Schlegel}, D.~J., {Eisenstein}, D.~J., \&
  {Brinkmann}, J. 2004, \aj, 127, 1860

\bibitem[{{de Boor}(1977)}]{deboor_bsplines}
{de Boor}, C. 1977, SIAM Journal on Numerical Analysis, 14, 441

\bibitem[{{Gavazzi} {et~al.}(2007){Gavazzi}, {Treu}, {Rhodes}, {Koopmans},
  {Bolton}, {Burles}, {Massey}, \& {Moustakas}}]{slacs4}
{Gavazzi}, R., {Treu}, T., {Rhodes}, J.~D., {Koopmans}, L.~V., {Bolton}, A.~S.,
  {Burles}, S., {Massey}, R., \& {Moustakas}, L.~A. 2007, \apj,
  \textit{(submitted) (astro-ph/0701589)}

\bibitem[{{Hewett} {et~al.}(1985){Hewett}, {Irwin}, {Bunclark}, {Bridgeland},
  {Kibblewhite}, {He}, \& {Smith}}]{hewett_prism}
{Hewett}, P.~C., {Irwin}, M.~J., {Bunclark}, P., {Bridgeland}, M.~T.~.,
  {Kibblewhite}, E.~J., {He}, X.~T., \& {Smith}, M.~G. 1985, \mnras, 213, 971

\bibitem[{{Hook} {et~al.}(2003)}]{hook_gmos}
{Hook}, I., {et~al.} 2003, in Instrument Design and Performance for
  Optical/Infrared Ground-based Telescop es. Edited by Iye, Masanori; Moorwood,
  Alan F. M. Proceedings of the SPIE, Volume 484 1, pp. 1645-1656 (2003).,
  1645--1656

\bibitem[{{Horne}(1986)}]{horne_optimal}
{Horne}, K. 1986, \pasp, 98, 609

\bibitem[{{Kassiola} \& {Kovner}(1993)}]{kassiola_kovner}
{Kassiola}, A., \& {Kovner}, I. 1993, \apj, 417, 450

\bibitem[{{Keeton} \& {Kochanek}(1998)}]{kk_spiral}
{Keeton}, C.~R., \& {Kochanek}, C.~S. 1998, \apj, 495, 157

\bibitem[{{Kelson}(2003)}]{kelson_skysub}
{Kelson}, D.~D. 2003, \pasp, 115, 688

\bibitem[{{Koopmans} {et~al.}(2006){Koopmans}, {Treu}, {Bolton}, {Burles}, \&
  {Moustakas}}]{slacs3}
{Koopmans}, L.~V.~E., {Treu}, T., {Bolton}, A.~S., {Burles}, S., \&
  {Moustakas}, L.~A. 2006, \apj, 649, 599

\bibitem[{{Kormann} {et~al.}(1994){Kormann}, {Schneider}, \&
  {Bartelmann}}]{kormann_sie}
{Kormann}, R., {Schneider}, P., \& {Bartelmann}, M. 1994, \aap, 284, 285

\bibitem[{{Lamer} {et~al.}(2006){Lamer}, {Schwope}, {Wisotzki}, \&
  {Christensen}}]{lamer_06}
{Lamer}, G., {Schwope}, A., {Wisotzki}, L., \& {Christensen}, L. 2006, \aap,
  454, 493

\bibitem[{{Mediavilla} {et~al.}(1998){Mediavilla}, {Arribas}, {del Burgo},
  {Oscoz}, {Serra-Ricart}, {Alcalde}, {Falco}, {Goicoechea}, {Garcia-Lorenzo},
  \& {Buitrago}}]{mediavilla_98}
{Mediavilla}, E., {Arribas}, S., {del Burgo}, C., {Oscoz}, A., {Serra-Ricart},
  M., {Alcalde}, D., {Falco}, E.~E., {Goicoechea}, L.~J., {Garcia-Lorenzo}, B.,
  \& {Buitrago}, J. 1998, \apjl, 503, L27+

\bibitem[{{Metcalf} {et~al.}(2004){Metcalf}, {Moustakas}, {Bunker}, \&
  {Parry}}]{met_mous_04}
{Metcalf}, R.~B., {Moustakas}, L.~A., {Bunker}, A.~J., \& {Parry}, I.~R. 2004,
  \apj, 607, 43

\bibitem[{{Motta} {et~al.}(2004){Motta}, {Mediavilla}, {Mu{\~n}oz}, \&
  {Falco}}]{motta_04}
{Motta}, V., {Mediavilla}, E., {Mu{\~n}oz}, J.~A., \& {Falco}, E. 2004, \apj,
  613, 86

\bibitem[{{Murray} {et~al.}(2003)}]{murray_gmos_ifu}
{Murray}, G.~J., {et~al.} 2003, in Instrument Design and Performance for
  Optical/Infrared Ground-based Telescop es. Edited by Iye, Masanori; Moorwood,
  Alan F. M. Proceedings of the SPIE, Volume 484 1, pp. 1750-1759 (2003).,
  1750--1759

\bibitem[{{Schlegel} {et~al.}(1998){Schlegel}, {Finkbeiner}, \&
  {Davis}}]{sfd_dust}
{Schlegel}, D.~J., {Finkbeiner}, D.~P., \& {Davis}, M. 1998, \apj, 500, 525

\bibitem[{{Schmoll} {et~al.}(2004){Schmoll}, {Dodsworth}, {Content}, \&
  {Allington-Smith}}]{schmoll_imacs_ifu}
{Schmoll}, J., {Dodsworth}, G.~N., {Content}, R., \& {Allington-Smith}, J.~R.
  2004, in Ground-based Instrumentation for Astronomy. Edited by Alan F. M.
  Moorwood and Iye Masanori. Proceedings of the SPIE, Volume 5492, pp. 624-633
  (2004)., ed. A.~F.~M. {Moorwood} \& M.~{Iye}, 624--633

\bibitem[{{Sugai} {et~al.}(2007){Sugai}, {Kawai}, {Shimono}, {Hattori},
  {Kosugi}, {Kashikawa}, {Inoue}, \& {Chiba}}]{sugai_07}
{Sugai}, H., {Kawai}, A., {Shimono}, A., {Hattori}, T., {Kosugi}, G.,
  {Kashikawa}, N., {Inoue}, K.~T., \& {Chiba}, M. 2007, \apj, \textsl{in press
  (astro-ph/0702392)}

\bibitem[{{Swinbank} {et~al.}(2006){Swinbank}, {Bower}, {Smith}, {Smail},
  {Kneib}, {Ellis}, {Stark}, \& {Bunker}}]{swinbank_06}
{Swinbank}, A.~M., {Bower}, R.~G., {Smith}, G.~P., {Smail}, I., {Kneib}, J.-P.,
  {Ellis}, R.~S., {Stark}, D.~P., \& {Bunker}, A.~J. 2006, \mnras, 368, 1631

\bibitem[{{Swinbank} {et~al.}(2007){Swinbank}, {Bower}, {Smith}, {Wilman},
  {Smail}, {Ellis}, {Morris}, \& {Kneib}}]{swinbank_07}
{Swinbank}, A.~M., {Bower}, R.~G., {Smith}, G.~P., {Wilman}, R.~J., {Smail},
  I., {Ellis}, R.~S., {Morris}, S.~L., \& {Kneib}, J.-P. 2007, \mnras,
  \textit{(in press)}, 72

\bibitem[{{Swinbank} {et~al.}(2003){Swinbank}, {Smith}, {Bower}, {Bunker},
  {Smail}, {Ellis}, {Smith}, {Kneib}, {Sullivan}, \&
  {Allington-Smith}}]{swinbank_03}
{Swinbank}, A.~M., {Smith}, J., {Bower}, R.~G., {Bunker}, A., {Smail}, I.,
  {Ellis}, R.~S., {Smith}, G.~P., {Kneib}, J.-P., {Sullivan}, M., \&
  {Allington-Smith}, J. 2003, \apj, 598, 162

\bibitem[{{Treu} {et~al.}(2006){Treu}, {Koopmans}, {Bolton}, {Burles}, \&
  {Moustakas}}]{slacs2}
{Treu}, T., {Koopmans}, L.~V., {Bolton}, A.~S., {Burles}, S., \& {Moustakas},
  L.~A. 2006, \apj, 640, 662

\bibitem[{{Wayth} {et~al.}(2005){Wayth}, {O'Dowd}, \& {Webster}}]{wayth_05}
{Wayth}, R.~B., {O'Dowd}, M., \& {Webster}, R.~L. 2005, \mnras, 359, 561

\bibitem[{{Wisotzki} {et~al.}(2003){Wisotzki}, {Becker}, {Christensen},
  {Helms}, {Jahnke}, {Kelz}, {Roth}, \& {Sanchez}}]{wisotzki_03}
{Wisotzki}, L., {Becker}, T., {Christensen}, L., {Helms}, A., {Jahnke}, K.,
  {Kelz}, A., {Roth}, M.~M., \& {Sanchez}, S.~F. 2003, \aap, 408, 455

\bibitem[{{York} {et~al.}(2000)}]{york_sdss}
{York}, D.~G., {et~al.} 2000, \aj, 120, 1579

\endrefs

\end{document}